\documentclass[preprint]{emulateapj}
\usepackage{epsf}
\def\gtsim {\lower .1ex\hbox{\rlap{\raise .6ex\hbox{\hskip .3ex
        {\ifmmode{\scriptscriptstyle >}\else
                {$\scriptscriptstyle >$}\fi}}}
        \kern -.4ex{\ifmmode{\scriptscriptstyle \sim}\else
                {$\scriptscriptstyle\sim$}\fi}}}
\newcommand{\etal}{{et al.~}}
\newcommand{\be}{\begin{equation}}
\newcommand{\ee}{\end{equation}}
\newcommand{\Mo}{\ {\rm M_\odot}}
\def\LCDM{$\Lambda$CDM}
\def\kpc{\ {\rm kpc}}
\def\E{{\cal E}}
\def\degrees{^\circ}

\begin{document}

\submitted{The Astrophysical Journal, accepted}
\vspace{1mm}
\slugcomment{{\it The Astrophysical Journal, accepted}}

\shortauthors{KAZANTZIDIS, ZENTNER, \& KRAVTSOV}
\lefthead{The Robustness of Dark Matter Density Profiles in Dissipationless Mergers}
\righthead{KAZANTZIDIS, ZENTNER, \& KRAVTSOV}

\title{The Robustness of Dark Matter Density Profiles in Dissipationless Mergers}

\author{Stelios Kazantzidis,\altaffilmark{1,2}
            Andrew R. Zentner,\altaffilmark{2}
            and Andrey V. Kravtsov\altaffilmark{2,3}}

\begin{abstract}
  We present a comprehensive series of dissipationless
  $N$-body simulations to investigate the evolution of density
  distribution in equal-mass mergers between dark matter (DM) halos
  and multicomponent galaxies. The DM halo models are constructed with
  various asymptotic power-law indices ranging from steep cusps
  to core-like profiles and the structural properties of the galaxy models
  are motivated by the {\LCDM} paradigm of structure formation.
  The adopted force resolution allows robust density profile estimates in
  the inner $\sim 1\%$ of the virial radii of the simulated systems. We
  demonstrate that the central slopes and overall shapes of the remnant
  density profiles are virtually identical to those of the initial systems
  suggesting that the remnants retain a remarkable memory of the
  density structure of their progenitors, despite the relaxation that
  accompanies merger activity. We also find that halo concentrations 
  remain approximately constant through hierarchical merging involving 
  identical systems and show that remnants contain significant fractions of
  their bound mass well beyond their formal virial radii. These conclusions hold 
  for a wide variety of initial asymptotic density slopes, orbital energies, and 
  encounter configurations, including sequences of consecutive merger events, 
  simultaneous mergers of several systems, and mergers of halos with 
  embedded cold baryonic components in the form of disks, spheroids, or both. As an 
  immediate consequence, the net effect of gas cooling, which contracts and steepens 
  the inner density profiles of DM halos, should be preserved through a period of 
  dissipationless major merging. Our results imply that the characteristic universal shape 
  of DM density profiles may be set early in the evolution of halos.
\end{abstract}


\altaffiltext{1}{Institute for Theoretical Physics, University of Z\"urich,
Winterthurerstrasse 190, CH-8057 Z\"urich, Switzerland; 
{\tt stelios@physik.unizh.ch}.}
\altaffiltext{2}{Kavli Institute for Cosmological Physics and 
        Department of Astronomy and Astrophysics, 
        The University of Chicago, 
        5640 South Ellis Avenue, Chicago, IL 60637 USA; 
        {\tt zentner@kicp.uchicago.edu}, {\tt andrey@oddjob.uchicago.edu}.}
\altaffiltext{3}{The Enrico Fermi Institute, 5640 South Ellis Ave., The University 
        of Chicago, Chicago, IL 60637 USA.}

\section{Introduction}
\label{sec:intro}

In the currently favored cold dark matter (CDM) paradigm of
hierarchical structure formation \citep[e.g.,][]{blumenthal_etal84},
galaxy assembly initiates with the gravitational collapse of overdense
regions into halos of dark matter (DM), the average density of which
exceeds that of baryonic matter by roughly six to one
\citep[e.g.,][]{spergel_etal03}. Bound in the potential wells of DM halos, 
baryons cool, condense, and form galaxies with a variety of properties
\citep[e.g.,][]{white_rees78,blumenthal_etal84}. Understanding the nonlinear 
gravitational collapse of the DM is the necessary first step in gaining insight
into the physical processes of galaxy formation.

During the past decade the mass distribution of virialized DM halos
has been the subject of extensive numerical investigations
(e.g., Dubinski \& Carlberg 1991; Navarro \etal 1996, hereafter NFW; 
Fukushige \& Makino 1997; Moore \etal 1999, hereafter M99; Klypin \etal 2001; 
Fukushige \& Makino 2001; Power \etal 2003; Tasitsiomi \etal 2004; 
Diemand \etal 2004; Fukushige \etal 2004). These studies have established
that the spherically-averaged density profiles of CDM halos 
can be described by an approximately ``universal'' formula 
over the entire range of mass and length scales that
these simulations can resolve. This universal density
profile is comprised of an inner steep cusp with 
$\rho(r) \propto r^{-\gamma}$, where $1 \lesssim \gamma \lesssim 1.5$, 
and a power-law decline $\rho(r) \propto r^{-3}$ in the outer parts. The origin of this 
simple form of the density distribution is not yet understood, 
but it has been conjectured that merging should play a noteworthy
role in determining DM halo profiles 
\citep[e.g.,][]{wechsler_etal02,zhao_etal03,loeb_peebles03,lu_etal05}.

Numerous authors have considered the effect of mergers on the
evolution of the density distribution of halos and galaxies using controlled 
numerical experiments \citep[e.g.,][]{white78,farouki_etal83,villumsen83,capelato_etal95,
syer_white98,barnes99,fulton_barnes01,boylan-kolchin_ma04,moore_etal04,
aceves_velazquez05}. These studies employed collisionless $N$-body merger simulations and
reached the conclusion that the density structure of the merger remnants is 
reminiscent of that of their progenitors. However, most of these 
investigations used idealized, non-rotating, spherical systems to
represent the halo and galaxy models and were restricted to a narrow
range of encounter orbital energies. Furthermore, the great majority of
these earlier studies did not explore the effect of hierarchical merging
on the evolution of the density distribution and with the notable exception of
\citet{aceves_velazquez05}, did not address the role of a baryonic
component and internal angular momentum in shaping the inner density
profiles of merger remnants. Yet, the gravitational field in the
central regions of galaxies is dominated by baryons and, in the
standard paradigm of structure formation, galaxies and galaxy halos in
the universe are the end result of a complex hierarchy of mergers.
Elucidating the effect of internal angular momentum on the evolution 
of density profiles is fundamental as previous studies have 
already indicated \citep[e.g.,][]{bullock_etal01b,ascasibar_etal04}.

Despite the inexorable increase in the dynamic range of current
numerical simulations, several important issues 
remain unresolved. For example, radiative cooling and condensation of baryons 
in the halo centers, which is a precursor of galaxy formation, 
steepens the density distribution of DM
\citep{zeldovich_etal80,barnes_white84,blumenthal_etal86,ryden_gunn87,gnedin_etal04}.
It is essential to ascertain whether late mergers, that take place
after the majority of baryonic cooling has occurred, can 
modify the net effect of dissipation on halo density 
profiles, as conjectured by \citet{loeb_peebles03} and \citet{gao_etal04}. 
In addition, it is worth asking if the density distributions of halos today 
are related to the density profiles of their progenitors and whether halos 
retain any memory of earlier epochs, as is expected from analytical
considerations \citep[e.g.,][]{mathur88,dehnen05}. Addressing these questions 
may aid in understanding the dependence of halo concentration on
halo mass \citep{navarro_etal97,bullock_etal01a,eke_etal01} 
and its relation to halo mass-accretion-rates and formation times 
\citep{wechsler_etal02,zhao_etal03}.

The complexity of halo formation in a cosmological context, with continuous mergers, 
accretion, and rapidly changing potential wells, hinders isolation of the mechanisms that 
drive the evolution of halo density profiles.
We have therefore decided to examine the evolution of density structure in
{\it controlled} mergers of DM halos and multicomponent 
galaxies using a large ensemble of dissipationless $N$-body simulations. 
We improve upon the work of previous authors by adopting self-consistent 
realizations for the DM halo models and employing realistic models for the progenitor 
galaxies motivated by the prevailing {\LCDM} paradigm of structure formation.
Our simulation set is carefully designed to allow an investigation of
a much larger parameter space than before and our primary goal is to
establish conclusively the degree to which the density profiles of remnants 
retain memory of the structures of their progenitors over a wide range of
inner density slopes, internal and orbital angular momenta, and in the
presence of cold baryonic components. 

We conduct numerical experiments that extend and expand upon those 
of earlier studies in at least two important aspects. 
For the first time, we perform a thorough study of the effect of 
a merger sequence on the density profile of DM halos in an 
attempt to mimic the hierarchical mass assembly that characterizes 
CDM-like cosmological models. Moreover, the impact of a stellar component 
on the density distribution of DM during mergers has not been addressed to date 
with detailed simulations of realistic progenitor galaxy models.  
As we illustrate below, the central density slopes of the
merging systems are notably robust and the overall shapes of
remnant density profiles are related to a remarkable degree to those of 
their progenitors. Our results firmly establish that equal-mass mergers will not suffice
to modify the central density slopes and overall shape profiles of the DM distribution.

The outline of this paper is as follows. In Section~\ref{sec:methods}, 
we present the halo and galaxy models used in this study and the 
numerical methods we employ to construct them. We also describe 
in detail our simulation techniques and conduct convergence tests and 
numerical relaxation experiments that illustrate the robustness of 
our results to numerical effects. Section~\ref{sec:results} contains 
results of the analysis 
applied to binary and multiple mergers of halos and multicomponent galaxies. 
Implications and extensions of these findings are discussed in 
Section~\ref{sec:discussion}. Finally, in Section~\ref{sec:summary}, 
we summarize our main results.

\section{Numerical Methods}
\label{sec:methods}

\subsection{Halo Models}
\label{sub:halomodels}

We consider DM halo models with density profiles that are described 
by the general form \citep{zhao96,kravtsov_etal98},
\be
   \rho(r)=\frac{\rho_{\rm s}} {(r/r_{\rm s})^\gamma [1+(r/r_{\rm s})^\alpha]^
   {(\beta-\gamma)/\alpha}}  \qquad\hbox{($r \leq r_{\rm vir}$)} \ ,
   \label{general_density}
\ee 
where $\gamma$ is the {\it asymptotic} inner slope of the profile,
$\beta$ is the outer slope, and $\alpha$ parametrizes the 
transition between the inner and outer profiles, with larger 
values of $\alpha$ corresponding to sharper transitions. 
Here, $\rho_{\rm s}$ is a characteristic inner
density and $r_{\rm s}$ denotes the scale radius defined as the 
distance from the center where the logarithmic slope is the average 
of the inner and outer slopes, 
$\rm d \ln \rho(r)/\rm d \ln r = -(\gamma + \beta)/2$. 
The logarithmic slope corresponding to the density profile of equation
(\ref{general_density}) is given by
\be
   \frac{\rm d \ln \rho (r)}{\rm d \ln r} = -\gamma -(\beta-\gamma)
   \frac{(r/ r_{\rm s})^\alpha}
   {\left[1+(r/r_{\rm s})^\alpha\right]} .
   \label{log_slope}
\ee
It is worth emphasizing that the choice of the functional form of the halo 
profiles is not unique. Several recent numerical studies have suggested that
the density distributions of DM halos are better represented by a
function with a logarithmic slope continuously varying as a power law
in radius \citep{stoehr_etal02,navarro_etal04,merritt_etal05,graham_etal05}.
However, the adopted profiles constitute a sufficiently realistic representation
of the density distribution of cosmological halos for the purposes of
this study.

We define the virial radius, $r_{\rm vir}$, as
the radius enclosing an average density equal 
to the virial overdensity, $\Delta_{\rm vir}$,
times the critical density for a flat universe, $\rho_{\rm crit}$.
Thus, the virial mass and virial radius are related through 
$M_{\rm vir} = (4/3) \pi \rho_{\rm crit} \Delta_{\rm vir} r_{\rm vir}^{3}$.
Throughout the paper we adopt the concordance {\LCDM} cosmological 
model ($\Omega_{\rm m}=0.3$, $\Omega_{\Lambda}=0.7$, $h=0.7$)
and assume $z=0$. The virial overdensity is then equal to 
$\Delta_{\rm vir} \simeq 103.5$ \citep[e.g.,][]{lacey_cole93,eke_etal98}. 
The general form of equation (\ref{general_density}) 
includes as special cases many of the density profiles 
commonly used to fit DM halos in cosmological simulations. For example, the NFW and 
\citet{moore_etal99} density profiles correspond to the parameters 
$[\alpha,\beta,\gamma]=[1,3,1]$ and 
$[\alpha, \beta, \gamma ] = [1.5, 3, 1.5]$, respectively. 

One of the primary goals of this study is to examine the evolution of the
inner density slopes during equal-mass dissipationless mergers. Hence, we fix the
outer logarithmic slope to $\beta = -3$, as demonstrated by
cosmological simulations \citep[e.g., NFW][]{moore_etal99,avila_reese_etal99}. 
However, density profiles with outer slopes $\beta\ge -3$ lead to cumulative mass
profiles that diverge as $r \rightarrow \infty$. In order to keep the
total mass finite, it is necessary to introduce a cutoff in the 
density profile. Truncating the profile sharply for
$r>r_{\rm vir}$, results in unphysical models, therefore, we implement
an exponential cutoff which sets in at the virial radius and turns off
the profile on a scale $r_{\rm decay}$. The truncation scale $r_{\rm
  decay}$ is a free parameter and controls the sharpness of the
transition. Explicitly, we model the density profiles of halos beyond
$r_{\rm vir}$ by
\be
   \rho(r)=\frac{\rho_{\rm s}} {c (1+c)^2} 
   \left(\frac{r}{r_{\rm vir}}\right)^{\kappa}
   \exp\left[-\frac{r-r_{\rm vir}}{r_{\rm decay}}\right]  
   (r>r_{\rm vir}) \ ,
   \label{exp_cutoff}
\ee
where $c\equiv r_{\rm vir}/r_{\rm s}$ is the concentration parameter.
Finally, in order to ensure a smooth transition between 
the profile interior to $r_{\rm vir}$ given by equation (\ref{general_density})
and the profile exterior to $r_{\rm vir}$ given by 
equation (\ref{exp_cutoff}), we require the logarithmic slope 
to be continuous at $r=r_{\rm vir}$. 
This implies 
\be 
\kappa=\frac{-\gamma - \beta c^\alpha}{1+c^\alpha} +
\frac{r_{\rm vir}}{r_{\rm decay}}.
\label{eps}
\ee 
Note that this procedure results in some additional bound 
mass beyond $r_{\rm vir}$, the precise amount of which 
depends upon the adopted model parameters. In the
models we consider here, the mass exterior to $r_{\rm vir}$
is typically about $\sim 10\%$ of the mass contained 
within the virial radius.

We construct $N$-body halo models for this study using the 
method described in \citet{kazantzidis_etal04a}, 
which is based on an explicit computation of the exact phase-space 
distribution function (DF). Under the assumptions of spherical symmetry and an 
isotropic velocity dispersion tensor, the DF depends only on the binding energy 
per unit mass $\E$. In this case, the DF is given by 
\be
   f(\E) =\frac{1}{\sqrt{8} \pi^2}\left[ \int_{0}^{\E}
   \frac{{\rm d}^2 \rho}{{\rm d} \psi^2} \frac{{\rm d}
   \psi}{\sqrt{\E-\psi}} + \frac{1}{\sqrt{\E}} \left (\frac{{\rm d} \rho}
   {{\rm d}\psi}\right)_{\psi=0}\right] 
   \label{edd_formula}
\ee   
\citep{eddington16}, where $\rho(r)$ and $\psi(r)$ are the density profile 
and {\it relative} gravitational potential corresponding 
to the halo model, respectively. 
The second term on the right-hand side in 
equation (\ref{edd_formula}) vanishes for any sensible 
behavior of $\psi(r)$ and $\rho(r)$ at
large distances. The ${\rm d}^2\rho/{\rm d} \psi^2$ factor in the
integrand would be difficult to deal 
with numerically, but it can be evaluated 
analytically using equations 
(\ref{general_density}) and (\ref{exp_cutoff}) for $\rho(r)$ 
to give an expression in which the only derivatives remaining are 
${\rm d}\psi/{\rm d} r$ and ${\rm d}^2\psi/{\rm d} r^2$.
Both of these can be written in terms 
of the density profile $\rho(r)$ and
its cumulative mass distribution $M(r)$,
\be
   \frac{{\rm d}^2 \rho}{{\rm d} \psi^2} =  \left (\frac{r^{2}}{G M}\right)^{2}
   \left[\frac{{\rm d}^2 \rho}{{\rm d} r^2} + \frac{{\rm d} \rho}{{\rm d} r} 
   \left (\frac{2}{r} - \frac{4\pi\rho\,r^{2}}{M} \right) \right].  
\ee
Thus, equation (\ref{edd_formula}) is reduced to a 
simple quadrature with no numerical differentiation 
required which is integrated numerically 
to obtain the DF for the anticipated values of the 
energy $\E$. A Monte Carlo realization of the $N$-body 
model is then readily generated by randomly 
sampling the particle positions and velocities 
from the DF. This method makes no assumptions 
about the functional form of the local velocity 
distribution, and therefore produces self-consistent 
equilibria that are ideal for studies of the 
detailed structure of collisionless systems. 

It is well established that cosmological halos exhibit 
departures from spherical symmetry and velocity isotropy
\citep[e.g.,][]{cole_lacey96}. Nevertheless, in \S~\ref{sub:multi_halo},
we analyze merger simulations between systems resulting from binary encounters
of the initial spherical and isotropic models. These merger remnants are 
known to reproduce the shape and velocity anisotropy trends found in cosmological
simulations \citep{moore_etal04}.

We model our halos with three density profiles specified by particular
choices of the parameters in equation (\ref{general_density}). The
first model follows the NFW density profile with $[\alpha, \beta,
\gamma]=[1,3,1]$, while the second and third models correspond to
profiles with steeper ($[\alpha, \beta, \gamma]=[2,3,1.7]$) and
shallower ($[\alpha, \beta, \gamma]=[2,3,0.2]$) inner slopes. In the
interest of brevity, throughout the remainder of the text, we shall
refer to these density profiles as the NFW, ``steep,'' and ``shallow''
profiles respectively. The motivation for using the shallow density 
profile is threefold. First, it is used to fit the inferred DM density 
distribution in some observed systems \citep[e.g.,][]{deblok_etal01}. 
Second, it corresponds to a regime applicable for 
warm DM (WDM) and decaying DM (DDM) models, which would introduce 
such a core in the profile \citep[e.g.,][]{hogan_dalcanton00,avila-reese_etal01}.
Third, it has been argued that the {\it extrapolated} density profiles of cosmological DM 
halos have cores and not cusps \citep{stoehr_etal02}. 
It is worthwhile to stress again 
that in our modeling $\gamma$ denotes the asymptotic slope 
of the density profile. Instead, the logarithmic slopes of the steep, NFW, and
shallow profiles at $1\%$ of $r_{\rm vir}$ are $\rm d \ln \rho /\rm d
\ln r \simeq -1.72$, $\simeq -1.21$, and $\simeq -0.24$, respectively.
Throughout the remainder of this paper we use the terms 
{\it inner} and {\it central} density slope to refer to the 
logarithmic slope of the spherically-averaged density profiles 
at the minimum resolved radii of the simulations.

Our choice of density slopes ensures that the employed
density profiles have significantly different shapes, a fact which
will be crucial for the interpretation of the results. Each of our
initial DM halos had a virial mass of $M_{\rm vir}=10^{12} \Mo$,
implying a virial radius of $r_{\rm vir} \simeq 256.7 \kpc$, and a
concentration of $c=12$ \citep{bullock_etal01a}, resulting in a scale
radius of $r_{\rm s} \simeq 21.4 \kpc$. The adopted value of $M_{\rm vir}$
serves merely practical purposes and does not imply anything
special about the particular choice of mass scale. Because we do not
consider non-gravitational processes such as gaseous dissipation, the
scale-free nature of gravity allows the rescaling of our models to any
system of units and hence the extension of our conclusions to
mergers between equal-mass systems of any mass scale.

\begin{figure*}[t]
\centerline{\epsfysize=3.5truein \epsffile{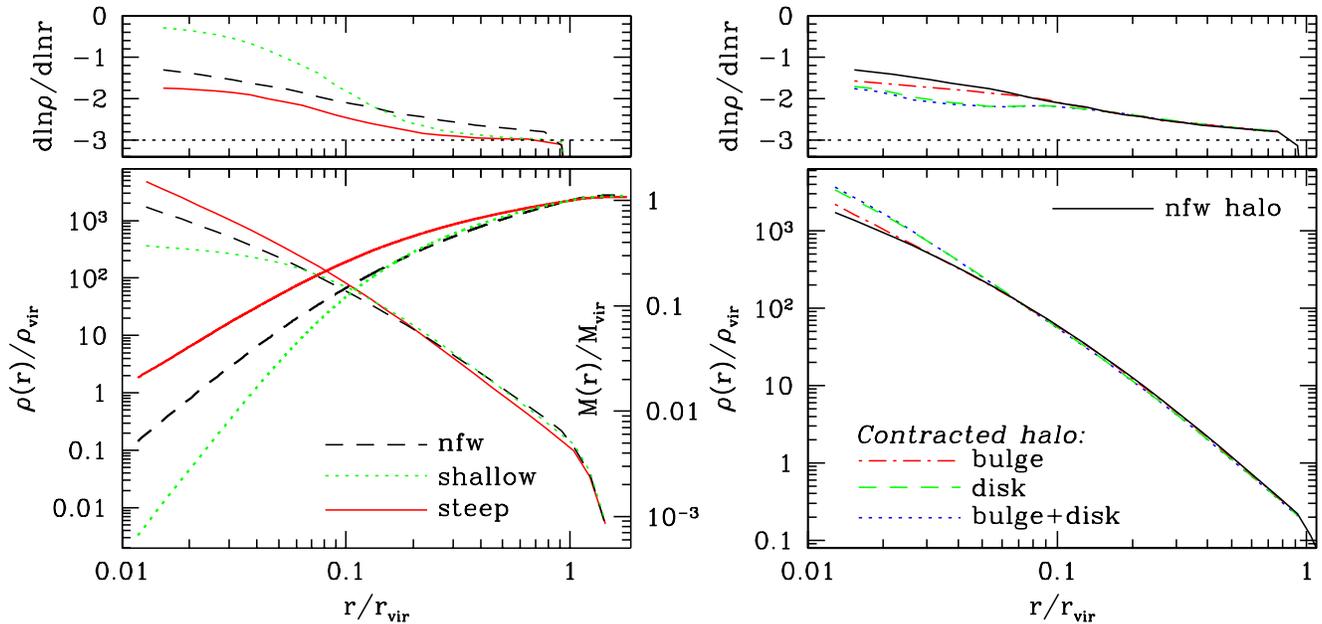}}
\caption{{\it Left:} Density, $\rho(r)$, ({\it bottom panel}) and 
  logarithmic slope profiles, $d \ln \rho(r)/d \ln r$, ({\it top panel}) for initial halo models as a 
  function of radius in units of the virial radius, $r_{\rm vir}$.
  Density is normalized to the virial density, $\rho_{\rm vir} \equiv \Delta_{\rm vir} \rho_{\rm crit}$. 
  The {\it solid}, {\it dashed}, and {\it dotted} 
  lines show results for the steep ($\gamma = 1.7$), NFW, 
  and shallow ($\gamma = 0.2$) profiles, respectively.  
  The initial density profiles are exponentially truncated 
  beyond $r_{\rm vir}$ because they correspond to 
  cumulative mass distributions that diverge at 
  large radii. The density slope profiles are computed by fitting 
  averaged density profiles locally at each radius. {\it Thick} lines 
  correspond to cumulative mass profiles for initial halo models normalized to the 
  virial mass, $M_{\rm vir}$ ({\it right axis}).
  {\it Right:} Density and logarithmic slope profiles of {\it DM} 
  after adiabatic contraction of an NFW density profile in response 
  to the growth of different stellar components.
  The profile before the baryonic infall corresponds to the {\it solid} curve.
  The {\it dot-dashed}, {\it dashed}, and {\it dotted} lines are the DM profile 
  after the growth of a bulge, disk, and a bulge+disk with parameters equal to those 
  of our fiducial disk galaxy model. The DM distribution responds 
  in degrees to the size of the total stellar component 
  included in the model and its central density profile approaches a power law with 
  slope $-2$ after the inclusion of both disk and bulge. 
\label{fig1}}
\end{figure*}

The left panel of Figure~\ref{fig1} shows the density, $\rho(r)$, ({\it bottom panel}) 
and logarithmic slope profiles, $d \ln \rho(r)/d \ln r$, ({\it top panel}) for all initial 
halo models as a function of radius in units of the virial radius, $r_{\rm vir}$. 
Density is given in units of the virial density, $\rho_{\rm vir} \equiv \Delta_{\rm vir} \rho_{\rm crit}$.
{\it Thick} lines correspond to the cumulative mass profiles for all initial halo models, $M(r)$, 
normalized to the virial mass, $M_{\rm vir}$. We computed density profiles 
from the force resolution ($2\epsilon$, where $\epsilon$ denotes the gravitational softening), 
outward defining the location of the most bound particle
as the center of the system. To reduce noise, density profiles were averaged over $25$ timesteps 
spanning a timescale equal to the crossing time at the virial radius of the system, 
$t_{\rm cross}(r_{\rm vir})=\sqrt{r_{\rm vir}^{3} / G M_{\rm vir}}$.
We derived logarithmic slope profiles by 
fitting the averaged density profile data locally about each radius. The accuracy of the density slope 
calculation was tested by comparing the results of the local fits against the exact analytical 
expressions for the initial models. The adopted fitting procedure yielded maximum
relative deviations of approximately $2\%$. This accuracy suffices for the purposes of 
our analysis. 

%

\subsection{Galaxy Models}
\label{sub:galmodels}

Disk galaxy models are constructed according to the procedure
described in \citet{hernquist93} and their structural parameters are
motivated by the currently favored galaxy formation paradigm in the
standard {\LCDM} model \citep{mo_etal98}. Each disk galaxy model
consists of a stellar disk embedded in a spherical and isotropic NFW
DM halo. The stellar disks follow an exponential distribution in cylindrical 
radius $R$ and their vertical structure is modeled by isothermal sheets
\be
   \rho_{\rm d}(R,z) = \frac{M_{\rm d}}{4\pi R_{\rm d}^2 z_{\rm d} } \exp\left(-\frac{R}{R_{\rm d}}\right)
   {\rm sech}^2\left(\frac{z}{z_{\rm d}}\right) \ ,
   \label{disk_density}
\ee
where $M_{\rm d}$, $R_{\rm d}$ and $z_{\rm d}$ denote the mass, 
radial scale length, and vertical scale height of the disk, respectively.
In our modeling, we parametrize the disk mass to be a fraction $m_{\rm d}$ of the 
halo virial mass, $M_{\rm d}=m_{\rm d}M_{\rm vir}$, and we specify the disk vertical scale 
height in units of the radial disk scale length.

While most of the models used in this study do not contain a bulge, 
we also consider more general galaxy models that include a compact 
bulge component. For simplicity, we ignore any rotation in the bulges 
and model them as non-rotating spheroids that follow the 
\citet{hernquist90} density profile
\be
   \rho_{\rm b}(r) = \frac{M_{\rm b}}{2\pi}\frac{a}{r(a+r)^3} \ ,
   \label{bulge_density}
\ee
where $M_{\rm b}$ is the bulge mass and $a$ its scale length. In analogy
to the treatment of the disk, we assume that the bulge mass is a fraction
$m_{\rm b}$ of the halo virial mass, $M_{\rm b}=m_{\rm b}M_{\rm vir}$, and 
parametrize the bulge scale length in units of the radial disk scale length, 
$a=f_{\rm b}R_{\rm d}$.

The DM halos are exponentially truncated beyond $r_{\rm vir}$ using
equation (\ref{exp_cutoff}) and have the same $M_{\rm vir}$ and $c$ as
the pure halo models. Once a set of cosmological parameters is
adopted, the virial quantities $M_{\rm vir}$ and $r_{\rm vir}$ are
uniquely determined by the halo circular velocity at the virial
radius, $V_{\rm vir}$ \citep{mo_etal98}. Furthermore, the DM halo
carries some net angular momentum specified by the
dimensionless spin parameter, $\lambda = J/G \sqrt{|E|/ M_{\rm vir}^5}$, 
where $J$ and $E$ are the total halo angular momentum
and energy, respectively. We follow \citet{springel_white99} and distribute the 
angular momentum of the DM halo by setting the halo streaming velocity to be a fixed 
fraction of the local total circular velocity. In our modeling, we assume that 
there is no exchange of angular momentum between the disk and the DM halo.  We 
also assume no angular momentum exchange between the disk and 
the bulge during galaxy formation. Thus, the specific angular 
momentum content of the disk is conserved.

Finally, the DM halos are adiabatically contracted in 
response to the growth of the collisionless stellar components under the 
assumptions of spherical symmetry, homologous contraction, circular DM particle orbits, 
and angular momentum conservation \citep{blumenthal_etal86}. The final DM distribution, 
$M_{\rm f}(r)$, can then be derived from the initial mass profiles of DM, $M_{\rm DM}(r)$, 
and baryons, $M_{\rm b}(r)$, and the final baryon distribution, $M_{\rm b}(r_{\rm f})$, 
according to 
\be
  \left[ M_{\rm DM}(r) + M_{\rm b}(r) \right] r =
  \left[ M_f(r) + M_{\rm b}(r_f) \right] r_f ,
  \label{equation:standard}
\ee
where $r_{\rm f}$ is the final radius of a DM particle. The right panel of 
Figure~\ref{fig1} displays the density and logarithmic slope profiles of {\it DM} after the
growth of different stellar components. For comparison, we also show the DM 
distribution of the initial NFW profile before the baryonic infall.
Adiabatic contraction modifies the DM density distribution significantly.
The inner profile approaches a power law with $\rho(r) \propto r^{-2}$ after including 
both the disk and bulge components.

For all disk galaxy models considered here, we adopt a halo spin
parameter equal to $\lambda=0.031$, which is close to the mean value
of spins measured in cosmological $N$-body simulations
\citep[e.g.,][]{bullock_etal01b}, a disk mass fraction of $m_{\rm d}=0.04$, 
and a constant vertical scale height across the disk equal to
$z_{\rm d}=0.1R_{\rm d}$. When a bulge component is added, its
mass and scale radius are specified by $m_{\rm b}=0.008$ and 
$f_{\rm b}=0.2$, respectively. The radial disk scale length $R_{\rm d}$ is 
uniquely determined for a given set of parameters $M_{\rm vir}$, $c$, $\lambda$, 
$m_{\rm d}$, $m_{\rm b}$, and $f_{\rm b}$. 
The resulting radial disk scale lengths are either 
$R_{\rm d}=3.8 \kpc$ or $R_{\rm d}=3.5 \kpc$, depending on whether or not the 
galaxy models are bulgeless. All disk galaxy models are initially dynamically stable
against bar formation (the mean Toomre Q-parameter of the disk was
about $1.4$) and their structural parameters are in accord with dynamical 
mass model A1 for the Milky Way presented in \citet{klypin_etal02}.

Particle positions for each component are initialized according to the
analytic density profiles. For the particle velocities, we assume
that the velocity distribution at any point in space is sufficiently
well approximated by a multivariate Gaussian whose mean velocity and
velocity dispersion tensor are given by the solution of the Jeans'
equations at this point (see \citet{hernquist93} for a detailed
discussion of sampling velocities). A critical reader may note that
disk galaxy models initialized under this scheme are not formally
self-consistent. Most interesting galaxy and halo models have local
self-consistent velocity profiles that become strongly non-Gaussian, 
especially near their centers. Indeed, \citet{kazantzidis_etal04b} and
\citet{springel_etal05} performed numerical experiments of isolated,
spherically-symmetric DM halos and established that the central
density cusps relax rapidly to an inner slope significantly shallower 
than the slope that was initially intended (see \citet{kazantzidis_etal04b} and
\citet{kazantzidis_etal04c} for a full discussion regarding the
implications of such instabilities for the evolution of substructure
in CDM halos and comparisons of the observed satellite dynamics with
the predictions of CDM models). 

While this is strictly correct, we argue that the galaxy models used in this study are 
not seriously affected by the approximate scheme that we employed.  
The reason is that the growth of different stellar components forces the inner DM density
profile to approach a power law with slope $\gamma \sim -2$. 
As the central density cusp becomes closer to the 
$\rho(r) \propto r^{-2}$ profile of a
singular isothermal sphere, the local velocity distribution approaches 
a Gaussian. In this case, the accuracy with which we compute 
the velocity distribution function is significantly improved.  
In fact, here we exploit the adiabatic contraction of the DM to construct 
{\it nearly} self-consistent disk galaxy models that do not show any discernible 
density evolution when evolved in isolation. We note that the numerical experiments
of \citet{kazantzidis_etal04b} adopted pure halo models with an inner
cusp of $\rho(r) \propto r^{-1}$, whereas the simulations of
\citet{springel_etal05} considered compound galaxy models with a
Hernquist DM halo and no adiabatic contraction.

For completeness, we also considered two-component elliptical galaxies
comprised of a spherical non-rotating stellar distribution embedded in
a spherical and isotropic NFW DM halo. The initial stellar
distributions followed the Hernquist density profile. We constructed
$N$-body models of elliptical galaxies adopting a procedure similar to the one
described in Section~\ref{sub:halomodels} for the realization of self-consistent 
one-component halo models. Briefly, in order to embed spherical
stellar distributions that are in equilibrium inside spherical DM halos, we
simply need to calculate the DF of component $i$, $f_{i}(\E)$. This is
done by adding the appropriate potential term to equation~(\ref{edd_formula})
\be
   f_{i}(\E) =\frac{1}{\sqrt{8} \pi^2} \int_{0}^{\E}
   \frac{{\rm d}^2 \rho_{i}}{{\rm d} \Psi^2} \frac{{\rm d} \Psi}{\sqrt{\E-\Psi}} \ ,
   \label{edd_formula_elliptical}
\ee   
where $\rho_{i}$ is the density profile of component $i$ and 
$\Psi(r) = \psi_{\rm DM}(r)+\psi_{\rm stars}(r)$ is 
the total gravitational potential. 

Finally, the elliptical galaxy models are constructed as before 
with the particle positions and velocities initialized from the corresponding DF. 
Note that the density profile of the DM halo will be modified 
when the stellar distribution forms in its center and will no longer be described 
by equation~(\ref{general_density}). We explicitly take this into account when 
calculating the halo DF. Similar to the disk galaxy models, 
we computed the change in the halo 
density structure by assuming that it responds adiabatically to the growth of the stellar 
component and using the standard form of the adiabatic contraction model of 
\citet{blumenthal_etal86}.

Before adiabatic contraction, the DM halos of the elliptical galaxies had the same 
virial mass and concentration as the pure halo models, 
$M_{\rm vir}=10^{12} \Mo$ and $c=12$.
The mass of the stellar distribution is chosen to be 
$M_{\rm stars}=0.1 M_{\rm vir}$. In order to assign the 
remaining free parameter of the Hernquist profile, namely the scale length $a$,
we followed \citet{boylan-kolchin_etal05} and used 
the observed relation between stellar mass and effective 
radius $R_{\rm e}$ for early-type 
galaxies in the Sloan Digital Sky Survey data \citep{shen_etal03},
\be
   R_{\rm e} = 4.16 \left( \frac{M_{\rm stars}}{10^{11} \, M_{\odot}} \right) ^{0.56} \, \kpc .
   \label{equation:sloan}
\ee
For the Hernquist profile, the effective radius $R_{\rm e}$ 
is related to the scale length $a$ 
by $R_{\rm e} \approx 1.82a$, giving $a \approx 2.3\kpc$ 
for our main elliptical galaxy model.
In the following section, we present a detailed 
description of all of the merger simulations that 
we undertake in this study.

\begin{deluxetable*}{lcccc|cccc}
\tablecaption{Summary of Merger Simulations}
\tablewidth{0pt}
\tablehead{
\multicolumn{6}{c}{{\bf Initial Models}}&\multicolumn{1}{c}{{\bf Remnants}}\\
\hline
\hline\\[1mm]
\colhead{Run}&
\colhead{$\gamma_1$}&
\colhead{$\gamma_2$}&
\colhead{Orbit}  &
\colhead{$\vert \rm d \ln \rho / \rm d \ln r \vert$}  &
\colhead{$M_{\rm vir}$} &
\colhead{$r_{\rm vir}$} &
\colhead{$\vert \rm d \ln \rho / \rm d \ln r \vert$} &
\\
\colhead{} &
\colhead{} &
\colhead{} &
\colhead{} &
\colhead{} & 
\colhead{($10^{12} \Mo$)} & 
\colhead{(${\rm kpc}$)}& 
\\
\colhead{(1)} &
\colhead{(2)} &
\colhead{(3)} &
\colhead{(4)} &
\colhead{(5)} &
\colhead{(6)} &
\colhead{(7)} &
\colhead{(8)} 
\\
}
\startdata
\\
HRBp & 1 & 1 & Parabolic & 1.16 & 1.42  & 288.7 & 1.13 \\
Bp1 & 1 & 1 & Parabolic & 1.31 & 1.42 & 288.7 & 1.30 \\
Bp2 & 0.2 & 0.2 & Parabolic & 0.29 & 1.42  & 288.7  & 0.28 \\
Bp3 & 1.7 & 1.7 & Parabolic & 1.74 & 1.56  & 297.7 & 1.73 \\
hBp1 & 0.2 & 1 & Parabolic & 0.29, 1.31 & 1.42  & 288.8  & 1.23  \\
hBp2 & 0.2 & 1.7 & Parabolic & 0.29, 1.74 & 1.48  & 292.5 & 1.70 \\
Br1 & 1 & 1 & Radial & 1.31 & 1.62  & 301.5 & 1.28 \\
Br2 & 0.2 & 0.2 & Radial & 0.29 & 1.61  & 301.1 & 0.28 \\
Br3 & 1.7 & 1.7 & Radial & 1.74 & 1.68  & 305.1 & 1.73 \\
hBr1 & 0.2 & 1 & Radial & 0.29, 1.31  & 1.62  & 301.4 & 1.21 \\
hBr2 & 0.2 & 1.7 & Radial & 0.29, 1.74 & 1.62  & 301.6 & 1.70 \\
Bc1 & 1 & 1 & Circular & 1.31 & 1.54  & 296.5 & 1.28 \\
Bc2 & 0.2 & 0.2 & Circular & 0.29 & 1.57  & 298.4 & 0.28 \\
Bc3 & 1.7 & 1.7 & Circular & 1.74 & 1.55  & 297.2 & 1.73 \\
hBc1 & 0.2 & 1 & Circular & 0.29, 1.31 & 1.53  & 296.0 & 1.22 \\
hBc2 & 0.2 & 1.7 & Circular & 0.29, 1.74 & 1.52  & 295.2 & 1.70  \\
Qp1 & 0.2 & 0.2 & Parabolic & 0.29 & 3.24  & 380.0 & 0.27 \\
Qp2 & 1.7 & 1.7 & Parabolic & 1.74 & 3.31  & 382.8 & 1.72 \\
Qr1 & 0.2 & 0.2 & Radial & 0.29 & 3.15  & 376.4 & 0.27 \\
Qr2 & 1.7 & 1.7 & Radial & 1.74 & 3.26  & 380.8 & 1.72  \\
Sp1 & 1 & 1 & Parabolic & 1.31 & 1.94  & 320.5  & 1.28  \\
Sp2 & 1 & 1 & Parabolic & 1.31 & 2.61  & 353.6 & 1.27 \\
Sp3 & 0.2 & 0.2 & Parabolic & 0.29 & 1.80  & 312.1  & 0.26 \\
Sp4 & 0.2 & 0.2 & Parabolic & 0.29 & 2.21  & 334.7 & 0.25 \\
Sp5 & 1.7 & 1.7 & Parabolic & 1.74 & 2.35  & 341.5 & 1.72 \\
Sp6 & 1.7 & 1.7 & Parabolic & 1.74 & 3.09  & 373.9 & 1.71 \\
egBp & 1 & 1 & Parabolic & 2.03 & 1.30  & 280.5 & 2.01 \\
hbBp & 1 & 1 & Parabolic & 1.57 & 1.40 & 287.5 & 1.55 \\
dgBp1 (coplanar) & 1 & 1 & Parabolic & 1.70 & 1.38  & 286.1 & 1.69 \\
dgBp2 (inclined) & 1 & 1 & Parabolic & 1.70 & 1.39  & 286.6  & 1.69  \\
dgBr (inclined) & 1 & 1 & Radial & 1.70 & 1.54  &  296.7 & 1.69 \\
hbdBp (inclined) & 1 & 1 & Parabolic & 1.76 & 1.38 & 286.1 & 1.74 \\
\enddata
\label{table:model_parameters}
\tablecomments{\small Columns (1)-(5) refer to the properties of the 
initial models.  Columns (6)-(8) refer to the properties of the 
merger remnants.  The quantities in each column are as follows.
Col. (1): Abbreviations for the merger simulations, B = binary, 
Q = quadruple, S = sequential, p = parabolic, r = radial, c = circular,
h = hybrid, eg = elliptical galaxy, dg = disk galaxy consisting of a 
halo and a disk, hb = galaxy consisting of a halo and a bulge, 
hbd = galaxy consisting of a halo, a disk, and a bulge. 
For mergers between galaxies containing disks  
the orientation of the merging disks is included in 
parentheses.  All initial halo models have 
$N = 2 \times 10^5$ particles except for run HRBp which refers to the 
high-resolution binary, parabolic merger between NFW halos with 
$N = 2 \times 10^6$ particles per progenitor.
Col. (2), (3): Asymptotic inner density slopes.
Note that for galaxy models the asymptotic inner 
density slope of the initial {\it uncontracted} NFW halo model is given.
Col. (4): Initial orbital configuration.  
Col. (5): Logarithmic density slope at the {\it minimum} resolved radii. 
For mergers of identical 
progenitors only one number is shown while for 
the hybrid mergers of two different initial models, two slopes are given.  
For the galaxy models we only give the 
density slope of the DM component.
We estimate that our determination of slopes is accurate to within 
$\sim 2\%$ at radii that are well resolved.
Col. (6): Virial mass in units of $10^{12} \Mo$. 
Col. (7): Virial radius in $\kpc$.
Col. (8): Logarithmic slope at the innermost resolved radius. 
}
\end{deluxetable*}
%

\subsection{Description of Merger Simulations}
\label{sub:sims}

All numerical calculations discussed in this paper were carried out 
using PKDGRAV, a multi-stepping, parallel, tree $N$-body code 
\citep{stadel01}. PKDGRAV uses a spline softening length, such 
that the force is completely Keplerian at twice the quoted 
softening length, and multi-stepping based on the local 
acceleration of particles. We used an adaptive,
kick-drift-kick leapfrog integrator with individual 
particle time steps $\Delta t_{\rm i}$ chosen according to 
$\Delta t_{\rm i} \leq \eta (\epsilon_{\rm i}/\alpha_{\rm i})^{1/2}$, 
where $\epsilon_{\rm i}$ is the gravitational softening length 
of the particle, $\alpha_{\rm i}$ is the value of the local 
acceleration, and $\eta$ is a parameter that specifies the 
size of the individual timesteps and consequently the time 
accuracy of the integration. 

\subsubsection{Numerical Parameters}
\label{subsub:numer}

For all merger simulations between pure DM halos, 
we used $N = 2 \times 10^5$ particles and employed 
a gravitational softening length of $\epsilon = 1.5\kpc$. 
This choice enables us to resolve density profiles to 
$\sim 1\%$ of the virial radii of the simulated systems. 
The adopted force resolution is comparable to the mean particle 
separation within the region we want to resolve and is in accord 
with recent studies suggesting optimal scalings between the number of 
particles, $N$, and the minimum resolved radii 
\citep[e.g.,][]{power_etal03,reed_etal05}.
In \S~\ref{sec:tests}, we quantify the numerical effects 
of two-body relaxation on the central regions of our models and conduct 
convergence tests to ensure that the adopted mass and 
force resolution is adequate for the purposes of this study. The results of 
these tests confirm that our $N$-body simulations are robust to artificial 
numerical effects.

For binary mergers between multicomponent systems,
we used $N = 2 \times 10^5$ particles 
to represent the DM halo and $N = 2 \times10^4$ 
collisionless stellar particles for each stellar component. 
Gravitational forces for the stellar components were computed using a 
softening length of $\epsilon = 0.25 \kpc$ and the force resolution of 
the DM component was the same as in the halo-only mergers. 
To quantify the extent of artificial numerical effects on our findings, 
we compared the results of these merger simulations against the higher resolution 
calculations presented in the study of \citet{kazantzidis_etal05}. 
These authors employed galaxy models consisting of $10^6$ DM particles 
and $10^5$ particles in each stellar component and correspondingly 
smaller softening lengths. The comparison showed excellent agreement and in the 
remainder of the paper we will always present results for merger simulations 
with the standard mass and force resolution.

For all simulations, we set the base-timestep to be equal to $1\%$ of the dynamical 
time at the half-mass radius of the model and allowed the individual 
particle timesteps to be at most a factor of $2^{30}$ smaller. 
The time integration was performed with high enough 
accuracy such that the total energy was 
conserved to better than $0.1\%$ in all cases, 
which is adequate for the type of study that we undertake in 
this paper. As we have already stated, one of our main goals is to investigate the 
evolution of the central density slopes on small scales. 
The total energy contained in the inner regions of our models 
is a few tens of a percent of that of the entire system, so 
the energy conservation accuracy must be at least comparable 
to that in order to resolve meaningfully the dynamics 
of the region of interest. 

\subsubsection{Orbital Parameters}
\label{subsub:orbits}

Initial conditions for binary mergers were generated by building
pairs of halo or galaxy models and placing them at a distance equal to twice 
their virial radii. In the coordinate system chosen to describe the merger simulations,
the orbital plane coincides with the $x-y$ plane, and the center of mass
of the system coincides with the coordinate origin.
We simulated binary mergers of systems on parabolic orbits in agreement with cosmological
expectations \citep[e.g.,][]{khochfar_burkert05}. 
In order to examine the effect of encounter geometry 
and initial orbital energy on the density structure of 
the merger remnants we also considered 
circular orbits and radial orbits with zero 
orbital angular momenta \citep{moore_etal04}.

For the parabolic mergers, we set the initial 
models on orbits with pericentric distances 
of $20\%$ of the halo virial radii, a value that is typical of merging halos 
in cosmological simulations \citep[e.g.,][]{ghigna_etal98,zentner_etal05a,benson05}. 
The initial center of mass velocity of each pair was determined from the 
corresponding Keplerian orbit of two point masses. 
The trajectories of the merging systems deviate shortly 
after they begin to overlap as orbital energy is dissipated
and eventually follow closer orbits. In the case of the radial mergers, we set 
the initial relative velocities to be equal in magnitude 
to the velocities of point particles in a circular orbit 
about the common center of mass. In this way, both the 
radial and circular mergers have precisely the same total orbital energy. 

For binary mergers between disk galaxy 
models it is also necessary to choose a relative 
orientation for the disk components. We simulated
parabolic and radial mergers using both randomly-inclined 
and coplanar prograde (disk spin vectors parallel to the
orbital angular momentum vector) disk orientations. A subset of these 
binary merger simulations were discussed in \citet{kazantzidis_etal04c} 
where additional details can be found. Finally, we performed a 
single parabolic binary merger between elliptical galaxies and 
halo$+$bulge systems aimed at establishing the generality of our basic conclusions.

\begin{figure*}[t]
\centerline{\epsfysize=3.5truein \epsffile{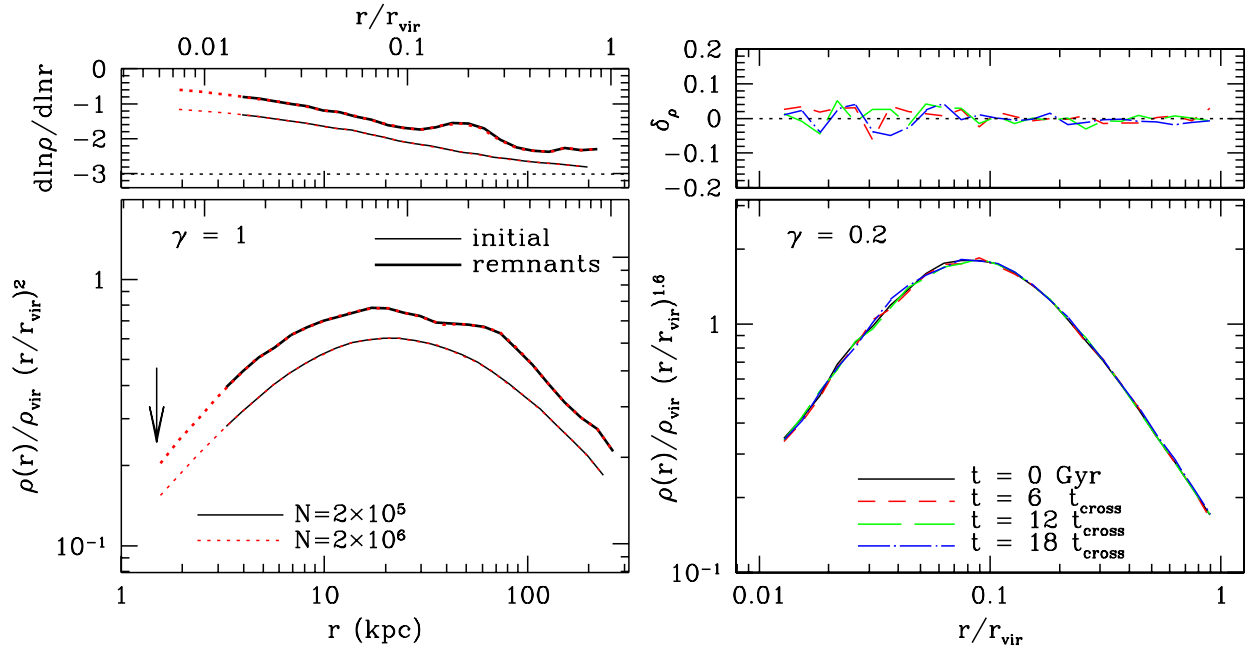}}
\caption{{\it Left:} Spherically-averaged density, $\rho(r)$, 
  ({\it bottom panel}) and logarithmic 
  slope profiles, $d \ln \rho(r)/d \ln r$, ({\it top panel})
  for two binary, parabolic merger simulations 
  between NFW halos (runs HRBp and Bp1). Each simulation has a different 
  force and mass resolution. In order to highlight any 
  differences at small radii, we plot the product $\rho(r) r^{2}$, 
  which is constant for isothermal distributions.
  The density profiles are plotted from the force resolution ($2\epsilon$) 
  outward and the number of particles, $N$, is indicated in the 
  {\it lower left-hand corner}. Radius is shown in both physical units ({\it bottom axis})
  and as a fraction of the virial radius 
  of the {\it initial} models, $r_{\rm vir}$ ({\it top axis}). 
  For clarity, the density slopes corresponding to the remnants
  ({\it thick lines}) have been shifted up by $0.5$~dex. 
  The downward arrow indicates the gravitational softening used in 
  the low-resolution simulation. The profiles are indistinguishable from the radius 
  corresponding to the force resolution of the lower resolution simulation 
  indicating that numerical convergence has been achieved.
  {\it Right:} Density profile as a function of time spanning 
  $18\,t_{\rm cross}(r_{\rm vir})$ for an 
  isolated halo following the shallow density profile. In order to 
  emphasize any differences at small radii, we plot the product 
  $\rho(r) r^{(\beta + \gamma)/2}$.  
  The maximum of this quantity indicates 
  the scale radius of the system. 
  {\it Top panel:} Relative density differences, 
  $\delta_{\rho} \equiv (\rho_{\rm in}-\rho_{\rm fi})/\rho_{\rm in}$,
  between the initial, $\rho_{\rm in}$, and final, $\rho_{\rm fi}$, 
  profiles for the same timescales. 
  Virtually no evolution in the density profile can be discerned over 
  the timescales of the simulation, indicating that our numerical 
  results are robust against two-body relaxation for evolution 
  on timescales smaller than $18\,t_{\rm cross}(r_{\rm vir})$.
\label{fig2}}
\end{figure*}

Besides conducting binary mergers, one of our additional goals was 
to assess the net effect of a {\it hierarchy} of mergers, 
which is a more appropriate characterization of the early 
evolution of halos in the standard paradigm of 
cosmological structure formation. Along these lines, 
we have simulated a sequence of binary, equal-mass 
mergers between identical halos for each of the shallow, 
NFW, and steep initial density profiles. 
For this ensemble of simulations we only utilized parabolic orbits. 
Each sequence of mergers consisted of three simulations. 
The first stage was simply the binary merger
between the initial halo models described above. 
In the second stage, we merged identical copies of the merger 
remnants from the first stage. We identified a time after which 
the central density profile of the remnant did not evolve significantly 
(changes of the order of $1-2\%$ were considered acceptable)
and removed all unbound particles. We underscore that the 
outer regions of remnants may evolve for much longer 
\citep[e.g.,][]{kazantzidis_etal04c} as density waves 
associated with the merger process may 
persist for an extended period of time after the 
cores of the halos coalesce and the merger would 
conventionally be deemed complete. 
Finally, we chose the initial relative orientation of the 
principal axes of the remnants randomly, 
placed the remnants at a relative distance equal to 
twice their virial radii, and set them on parabolic orbits. 
For the third stage, 
we repeated the previous process using as 
progenitor systems the remnants of the previous stage. 

We determined the bound mass in each of the remnants using 
the iterative scheme described in \citet{kazantzidis_etal04b}. 
In the rest frame of the most bound particle,
we calculated the binding energies of all other particles 
using the tree-based gravity calculation performed 
by PKDGRAV and we removed all particles with positive 
binding energy. We repeated this calculation 
of binding energies and subsequent removal of unbound particles 
until no more unbound particles were found. 
In practice this iterative procedure converges rapidly 
and ensures that the true, bound entity will be identified. 
This technique is essentially the same used in most 
group-finding routines, like the publicly-available 
SKID \citep{stadel01}, but has the advantages of using a 
tree structure for the potential computation, which 
requires $\mathcal{O}(N \log N)$ operations instead of 
$\mathcal{O}(N^2)$ for $N$ particles, and a parallel 
implementation for very large $N$. In this way, a much 
larger number of particles can be 
used in the calculation than would be possible with SKID at a 
fraction of the computational burden.

In addition, we considered the simultaneous merging of four 
systems. Initial conditions for these quadruple mergers were generated 
by building halo pairs identical to the ones used in the 
binary merger experiments. We considered only shallow and 
steep initial density profiles, as these are the extremes of 
the range of profiles that we studied with the binary merger 
experiments, and set the center of mass position and velocity 
of the first pair to be the same as in the binary merger 
experiments. For the second pair, we changed the signs of 
the center of mass coordinates of each halo in such a way that 
its orientation was rotated $90\degrees$ with respect to the 
first pair. We simulated quadruple mergers in which 
each pair was initially set on either a parabolic or a radial orbit. 
The trajectories of the merging systems will strongly 
deviate from a parabolic and radial orbit shortly after the beginning 
of the simulation, but here we are not interested in reproducing a 
particular orbital configuration. The setup of these experiments 
simply ensures that four halos that correspond to the steep and 
shallow density profiles will merge following orbits with very 
different orbital configurations and orbital energies. 
However, for convenience, we shall refer to these quadruple 
mergers as ``parabolic'' and ``radial''.

In addition to the DM-only halo mergers, we studied the merging of the
multicomponent galaxy models described in \S~\ref{sub:galmodels}. We
studied mergers of identical galaxies with stellar disks and bulges
embedded in adiabatically contracted DM halos as well as mergers of
galaxies with only stellar disks or bulges. For the halo$+$disk
mergers, we studied variations in both the initial orbital
configuration and the initial relative orientation of the stellar
disks. We considered mergers of disks randomly inclined with respect
to the orbital plane on both parabolic and radial initial orbits and
mergers of coplanar prograde disks on parabolic initial orbits.
For the mergers of halo$+$disk$+$bulge compound galaxy models the
stellar disks were randomly inclined and the initial orbit was
parabolic. Finally, for the halo$+$bulge and elliptical galaxy
mergers, parabolic was again the initial orbit of choice. 
Table~\ref{table:model_parameters} contains a summary of all 
merger simulations performed in this study and a list of parameters and 
variables for all initial models and remnants.

\subsection{Numerical Tests}
\label{sec:tests}

In order to minimize any concern that our results might be compromised
by artificial numerical effects, we performed experiments
varying the mass resolution by a factor of $10$ and scaling down the
softening lengths according to $\epsilon \propto N^{-1/3}$. Results
from one of these tests are displayed in the left panels of Figure~\ref{fig2}.  
Although the merger remnants exhibit 
significant departures from spherical symmetry, it is instructive to calculate 
spherically-averaged remnant density profiles.
Fig.~\ref{fig2} shows {\it average} density $\rho(r)$, 
and logarithmic slope profiles $d \ln \rho(r)/d \ln r$, 
for two parabolic, binary mergers between NFW halos 
at two different numerical resolutions (runs HRBp and Bp1).
The agreement in the density and logarithmic slope profiles 
of the remnants simulated at different resolutions indicates that 
numerical convergence has been achieved.

Two-body relaxation is a numerical artifact 
associated with the use of particles to sample phase-space in the collisionless limit.
The coarse-grained sampling of phase-space 
originating from the enormous difference between 
the mass of a particle in $N$-body simulations and that of 
DM particle candidates results in a mean potential 
that can be dominated by two-body interactions. 
As a result, two-body relaxation sets a limit on 
the region of an $N$-body system within which 
the numerical results can be trusted. 
All of our initial models have a central density slope 
shallower than that of an isothermal ($\rho \propto r^{-2}$) model, 
therefore energy transfer due to two-body 
relaxation would cause an expansion of the central region 
and a subsequent flattening of the inner density slope, 
potentially interfering with 
the interpretation of our results.

To thoroughly investigate the effects of two-body 
relaxation on our results, we performed a test 
simulation. We evolved a halo in isolation following the 
shallow density profile for a total elapsed time of 
$t_{\rm relax} = 18\,t_{\rm cross}(r_{\rm vir})$. 
For the particular model we studied 
this corresponds to $t_{\rm relax} \simeq 34$~Gyr. 
The results of this experiment are shown in the right panels of Figure~\ref{fig2}. 
These panels present the density profiles, $\rho(r)$, ({\it bottom panel}) and the relative 
density differences, $\delta_{\rho} = (\rho_{\rm in}-\rho_{\rm fi})/\rho_{\rm in}$, ({\it top panel})
between the initial $\rho_{\rm in}$, and final $\rho_{\rm fi}$, profiles as a function 
of time. The density differences are small (of the order of a few percent), 
indicating that discreteness effects associated with the finite number of 
particles do not cause this particular halo model to evolve away from its
equilibrium configuration over these timescales.  
We stress that this shallow halo model will be the 
{\it most} susceptible to two-body relaxation owing to the relatively small 
number of particles in its central region.  Therefore, we anticipate that 
all of our simulations should be unaffected 
by numerical relaxation for evolution on timescales 
that are smaller than $t_{\rm relax}$ above.

\section{Results}
\label{sec:results}

\subsection{Binary Halo Mergers}
\label{sub:binary_halo}

We begin with the binary merger simulations between
pure DM halos. In all parabolic and radial mergers, the systems merge in 
three to five orbits (between $\sim 5$ to $\sim 7$ Gyr), with the specific
value depending on the inner power-law indices 
of the systems and the initial orbital configurations of the
mergers.  These timescales are larger than those reported 
in earlier studies of binary equal-mass mergers 
\citep[e.g.,][]{barnes_hernquist96} due to the larger and more 
realistic pericentric distances adopted here.  
The merger timescale is set by the combination 
of dynamical friction, which dissipates the orbital 
energy of the dense halo cores, and mass loss processes 
\citep[e.g.,][]{taffoni_etal03}. Models with steeper inner density 
distributions merge faster due to the stronger gravitational drag exerted by 
the galaxy cores. The differences in merging times can be of the order of
$\sim 1$~Gyr or even larger between shallow and steep density profiles. Finally, 
for circular orbits the merger process takes substantially longer time to complete 
owing to the much slower rate at which orbital energy is dissipated.

\begin{figure}[t]
\centerline{\epsfysize=7.85truein \epsffile{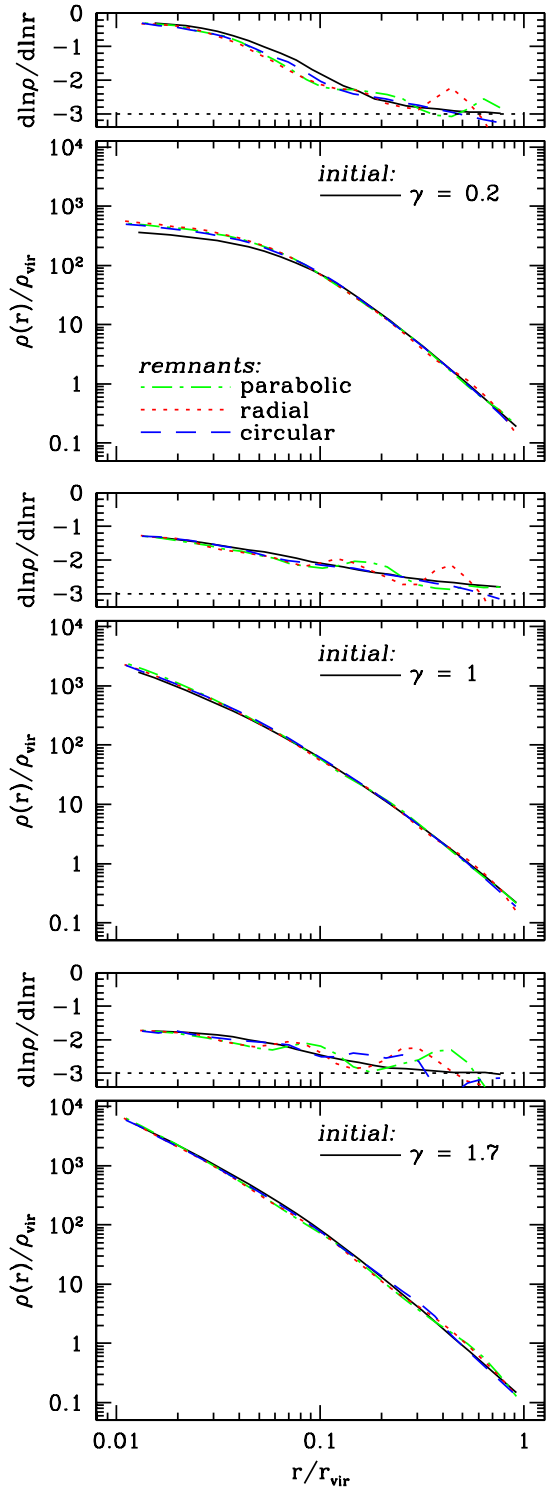}}
\caption{Spherically-averaged density and logarithmic slope profiles 
  for initial systems ({\it solid lines}) and remnants in binary merger simulations 
  between identical DM halo models (runs Bp1-Bp3, Br1-Br3 and Bc1-Bc3). The profiles are 
  plotted as a function of radius in units of the virial radius and each of the remnant profiles is 
  normalized to its own $r_{\rm vir}$. The asymptotic density power-law index, $\gamma$, 
  of the initial profiles is indicated in the {\it upper right-hand corner} 
  of each bottom panel. 
  The {\it dot-dashed, dotted,} and {\it dashed} lines 
  correspond to remnants in the parabolic, 
  radial, and circular initial orbital configurations, respectively.
\label{fig3}}
\end{figure}
\begin{figure*}[t]
\centerline{\epsfysize=2.9truein \epsffile{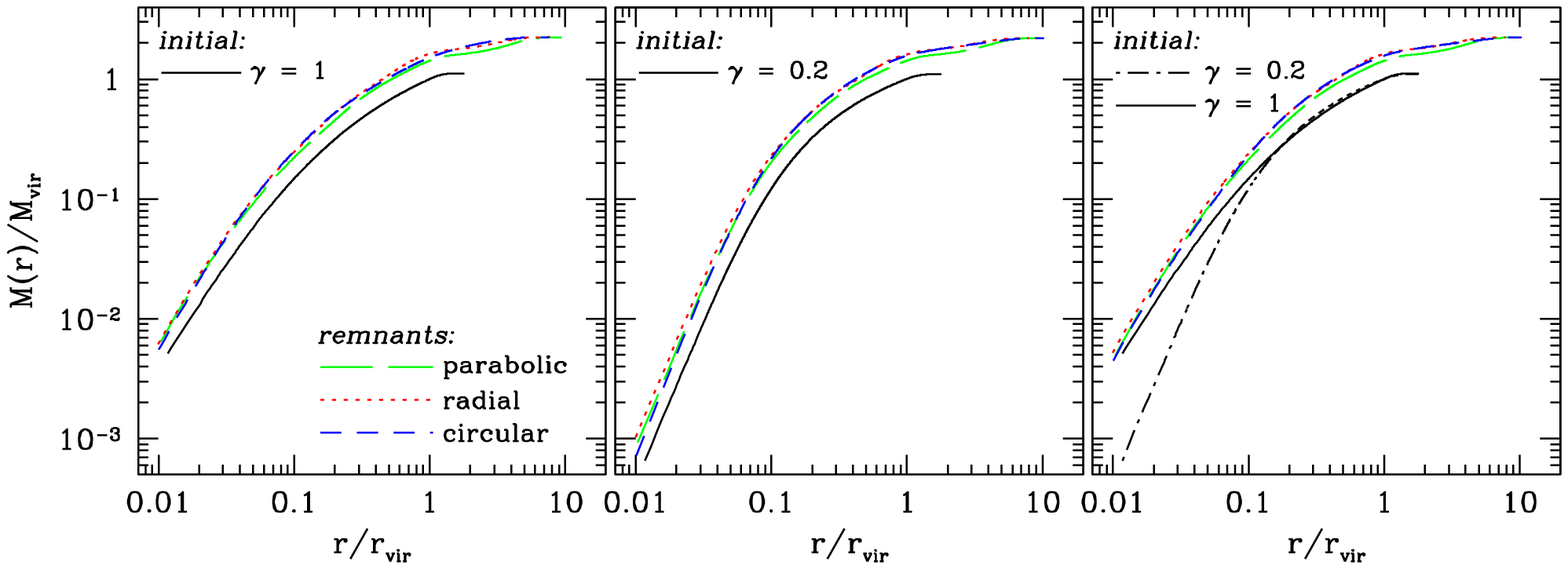}}
\caption{Cumulative mass profiles for remnants in binary
  merger simulations between halos as a function of radius in units of
  $r_{\rm vir}$. All profiles are normalized to the virial mass of the
  {\it initial} halo models, $M_{\rm vir}$. The two leftmost panels
  correspond to mergers between identical halo models with the initial
  value of $\gamma$ indicated in the {\it upper left corners} (runs Bp1, Bp2, Br1, Br2, Bc1 and Bc2). 
  The cumulative mass profiles of the initial halos are shown by the 
  {\it solid} lines. The right panel corresponds to hybrid mergers
  between an NFW halo ({\it solid} line) and a halo following the shallow 
  density profile ({\it dashed} line) (runs hBp1, hBr1 and hBc1). The different remnants correspond to different initial
  orbital configurations. The mass distribution of merger remnants extends well beyond the 
  virial radius (note that $\gtrsim 90\%$ of mass in the initial models is contained 
  within $r_{\rm vir}$).
\label{fig4}}
\end{figure*}

The first results that we present are those of the binary mergers involving identical halo 
models (runs Bp1-Bp3, Br1-Br3 and Bc1-Bc3). In Figure~\ref{fig3}, we show the 
spherically-averaged density, $\rho(r)$, 
({\it bottom panels}) and logarithmic slope profiles, $d \ln \rho(r)/d \ln r$, 
({\it top panels})
of the remnants for all nine identical-halo mergers along with 
the initial density profiles of each model. 
Note that each of the remnant profiles is normalized 
to its own virial radius, $r_{\rm vir}$,
as defined in \S~\ref{sub:halomodels}. Figure~\ref{fig3} demonstrates that both
the inner power-law indices and overall shape of the remnant density distributions
are strikingly similar to those of the progenitor systems. We stress that by the shape of the 
density profile, we refer to the radial dependence of its logarithmic slope as a function of the scaled 
radius $r/r_{\rm virial}$. This basic result holds for all initial density profiles and orbital energies that 
we considered.

\begin{figure*}[t]
\centerline{\epsfysize=4truein \epsffile{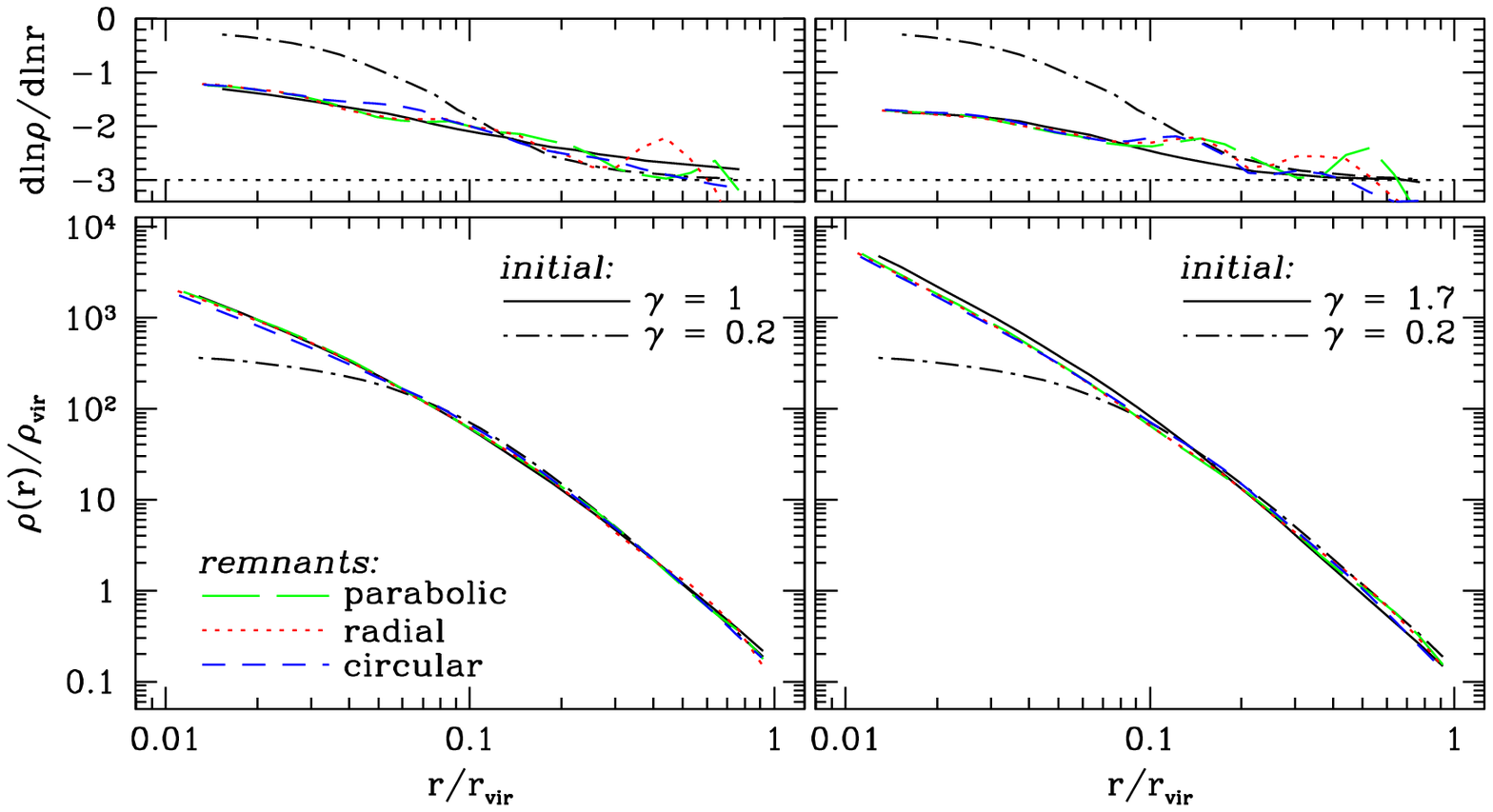}}
\caption{Density and logarithmic slope profiles as in Figure~\ref{fig3}, but for 
  binary mergers between halos with {\it different} initial central density power-law 
  indices $\gamma$ (runs hBp1, hBr1, hBc1, hBp2, hBr2, and hBc2).
  The initial asymptotic density slopes are indicated in the {\it upper right corners} 
  of the bottom panels. The steeper of the two initial profiles in each case
  is shown by the {\it solid} lines while the shallower of 
  the two initial profiles is depicted by the 
  {\it dot-dashed} lines. 
\label{fig5}}
\end{figure*}

Figure~\ref{fig4} presents cumulative mass profiles, $M(r)$, for the same set of
binary merger simulations. The left and middle panels show mass profiles for
mergers of identical initial halo models (runs Bp1, Bp2, Br1, Br2, Bc1 and Bc2), 
while the right panel corresponds to the ``hybrid'' halo mergers (runs hBp1, hBr1 and hBc1) 
which we will discuss in more detail below. Clearly, the cumulative measure reveals comparably
less about the details of the inner density profiles, but this plot
does display an interesting fact. The merger remnants contain
significant fractions of their bound mass, $\sim 25-35\%$ as opposed
to $\sim 10\%$ for the initial models, outside of what we formally
identify as their virial radii. Parabolic orbits result in a larger fraction 
of particles outside $r_{\rm vir}$ compared to bound orbits
owing to the fact that there is a larger amount of energy available
to distribute in the former (see also Table~\ref{table:model_parameters}).
These particle fractions are consistent with studies of 
halos in cosmological simulations that find that 
equilibrium density profiles extend well beyond halo virial radii 
\citep{prada_etal05}. 

One of the practical consequences of this result concerns the
ambiguous definition of the virial mass and the evolution of the
virial mass in analytic models for galaxy and halo formation
\citep[e.g.,][]{somerville_primack99,zentner_bullock03, benson_etal03,taylor_babul04,
zentner_etal05a}. The results reported in Figure~\ref{fig4} 
illustrate that a non-negligible 
amount of DM in the initial halo models is heated outside 
of the virial radius of the remnant. 
In fact, a few percent of the initial mass of the halos (between $\sim
1-3 \%$) becomes unbound in typical halo mergers.

Analytic models typically adopt a particular definition of halo virial
mass and then assume that the virial mass is strictly additive during
mergers. In contrast, the mass profiles in
Figure~\ref{fig4} show that, for a common definition of the virial
radius, the virial mass of the remnant is not simply the sum of the
virial masses of the progenitors. This suggests that a more
sophisticated approach is warranted in semi-analytic calculations
that use extended Press-Schechter merger trees to follow the mass
assembly histories of halos. 

Following up on the findings reported in Figs.~\ref{fig3} and \ref{fig4},
are the results of the hybrid binary merger simulations in 
which we merged DM halo-only models with substantially different values of 
the asymptotic density power-law index, $\gamma$ (runs hBp1, hBp2, hBr1, hBr2
hBc1 and hBc2). The density profiles that
result from the hybrid merger simulations are shown in
Figure~\ref{fig5}. This Figure reveals that, in equal-mass mergers
between halos with different power-law central density indices, the
remnant has a profile with an inner slope close to that of the
steepest of the progenitors \citep[see also][]{fulton_barnes01,
boylan-kolchin_ma04}. Similar to the DM 
halo-only mergers of identical initial models, this result is
independent of the precise value of the initial inner power-law slopes 
and the initial orbital energy of the progenitor halos. In the 
hybrid mergers, the net result appears to be that the inner core 
of the steepest of the two initial systems survives essentially intact in 
the remnant, an effect noted previously by \citet{boylan-kolchin_ma04} 
for mergers between NFW halos and halos with an inner core of constant 
density.

It is interesting to stress that the isodensity contours in the inner
region $r \lesssim 0.1 r_{\rm vir}$ of a hybrid merger remnant are
nearly spherical (both $b/a$ and $c/a$, where $a \ge b \ge c$ are the
principle axis ratios, are larger than $0.8$). Strongly prolate
remnants are emblematic of parabolic and radial mergers similar to the
ones simulated here \citep[e.g.,][]{moore_etal04,kazantzidis_etal04c}. 
The almost spherical inner regions of the hybrid merger remnants lends
support to the idea that the steepest inner spherical core survives,
almost unaltered, in the center of the final system.

\begin{figure*}[t]
\centerline{\epsfysize=3.5truein \epsffile{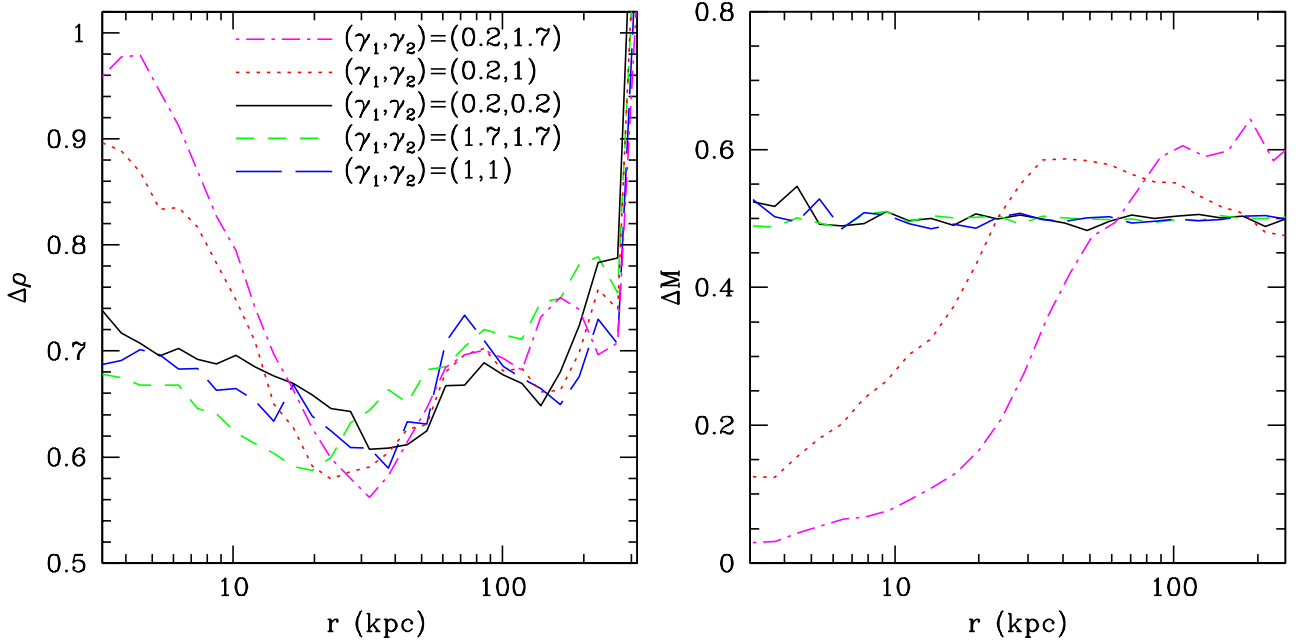}}
\caption{{\it Left:} Density fraction, $\Delta \rho \equiv \rho_{\rm rem}/2\rho_{\rm in}$,
  for all binary, parabolic mergers between halos as a function of radius
  in absolute units. Here $\rho_{\rm rem}$ and $\rho_{\rm in}$ denote
  the densities of remnants and initial models, respectively. In
  mergers between systems with different asymptotic central density
  power-law indices, $\gamma_{1}$ and $\gamma_{2}$, we define the
  fiducial quantity $2\rho_{\rm in} \equiv \rho_{\rm in}^{\gamma_{1}}
  + \rho_{\rm in}^{\gamma_{2}}$, where $\rho_{\rm in}^{\gamma_{1}}$
  and $\rho_{\rm in}^{\gamma_{2}}$ denote the initial density
  profiles of the two systems and $\gamma_{1}$ corresponds to the smaller 
  asymptotic power-law slope.
  {\it Right:} Relative contribution of mass in the remnants of binary, 
  parabolic mergers as a function of radius in physical units. $\Delta M$ 
  is defined as $\Delta M \equiv M_{\gamma_{1}}(r)/M_{\rm rem}(r)$, where 
  $M_{\rm rem}(r)$ is the total mass of the remnant and $M_{\gamma_{1}}(r)$ 
  denotes the mass of the progenitor halo with an asymptotic power-law index 
  $\gamma_{1}$ in the remnant. Line types are as in the left panel.
\label{fig6}}
\end{figure*}

In Figure~\ref{fig6}, we address the relative contribution of the
progenitor halos to the remnant density profiles in {\it parabolic} mergers. 
In the left panel we show the quantity 
$\Delta \rho \equiv \rho_{\rm rem}(r)/2\rho_{\rm in}(r)$, where $\rho_{\rm rem}(r)$ and
$\rho_{\rm in}(r)$ denote the densities of remnants and initial
models, respectively. In the hybrid mergers, we have defined the
fiducial quantity $2\rho_{\rm in}(r) \equiv \rho_{\rm
  in}^{\gamma_{1}}(r) + \rho_{\rm in}^{\gamma_{2}}(r)$, where
$\rho_{\rm in}^{\gamma_{1}}(r)$ and $\rho_{\rm in}^{\gamma_{2}}(r)$
correspond to the initial density profiles of the two systems and $\gamma_{1} < \gamma_{2}$.
In the right panel, we show the relative contribution of the progenitor halos
to the cumulative mass profiles of the remnants, $\Delta M \equiv
M_{\gamma_{1}}(r)/M_{\rm rem}(r)$. Here $M_{\rm rem}(r)$ is the
total mass of the remnant and $M_{\gamma_{1}}(r)$ denotes the mass in
the remnant that belongs to the progenitor halo with inner power-law
index $\gamma_{1}$.

As anticipated, in mergers of identical progenitor halos, the
contribution to the remnant from each halo is equal by symmetry, but
the variations in these quantities indicate the size of variations 
that are to be expected from numerical noise. For the hybrid encounters,
the mass from the steeper progenitor dominates the inner regions 
($r \lesssim 20 \kpc$ for the $[\gamma_1,\gamma_2] = [0.2,1.0]$ merger and
$r \lesssim 60 \kpc$ for the $[\gamma_1, \gamma_2] = [0.2,1.7]$
merger) of the final density profiles. For example, in the
$[\gamma_1,\gamma_2] = [0.2,1.0]$ merger, $80\%$ of the mass within
the inner $10 \kpc$ of the remnant originated from the NFW halo, while
in the $[\gamma_1, \gamma_2] = [0.2,1.7]$ merger, the steep halo
contributes fully $90\%$ of the remnant mass out to a distance of more
than $20 \kpc$ from the center of the remnant halo. 
Figure~\ref{fig6} also illustrates the fact that the resulting
remnant profile is not simply a sum of the two profiles of
the progenitors. The density of the remnant everywhere within
the virial radius is smaller than twice the density of the initial
models. This is because, as we noted above, mass is dynamically 
heated during the relaxation accompanying the merger and subsequently 
moves to larger radii. 

%

\subsection{Sequential and Multiple Halo Mergers}
\label{sub:multi_halo}

Next, we consider the merger sequences (runs Sp1-Sp6). In the sequential merger
simulations, identical copies of the remnants from the identical halo
mergers discussed in \S~\ref{sub:binary_halo} were collided with each
other after removing the small number ($\sim 1-3 \%$ in all cases) of
unbound particles. We initialized the mergers on parabolic orbits,
placed the systems at relative distances equal to twice their virial
radii and oriented the principal axes of the merger remnants randomly
with respect to each other. We repeated this process again to yield a
sequence of three merger remnants. We shall refer to the three levels
of mergers as ``level one,'' ``level two,'' and ``level three,''
respectively. The level one mergers refer simply to the parabolic
mergers discussed in \S~\ref{sub:binary_halo}. Note that the
pericentric distances for all merger levels are kept equal to $20\%$
of the halo virial radii. This choice is motivated by the results of
the binary merger simulations which demonstrated that the density structure
of the remnants is largely insensitive to the details of the encounter 
orbital energy.

\begin{figure*}[t]
\centerline{\epsfysize=4truein \epsffile{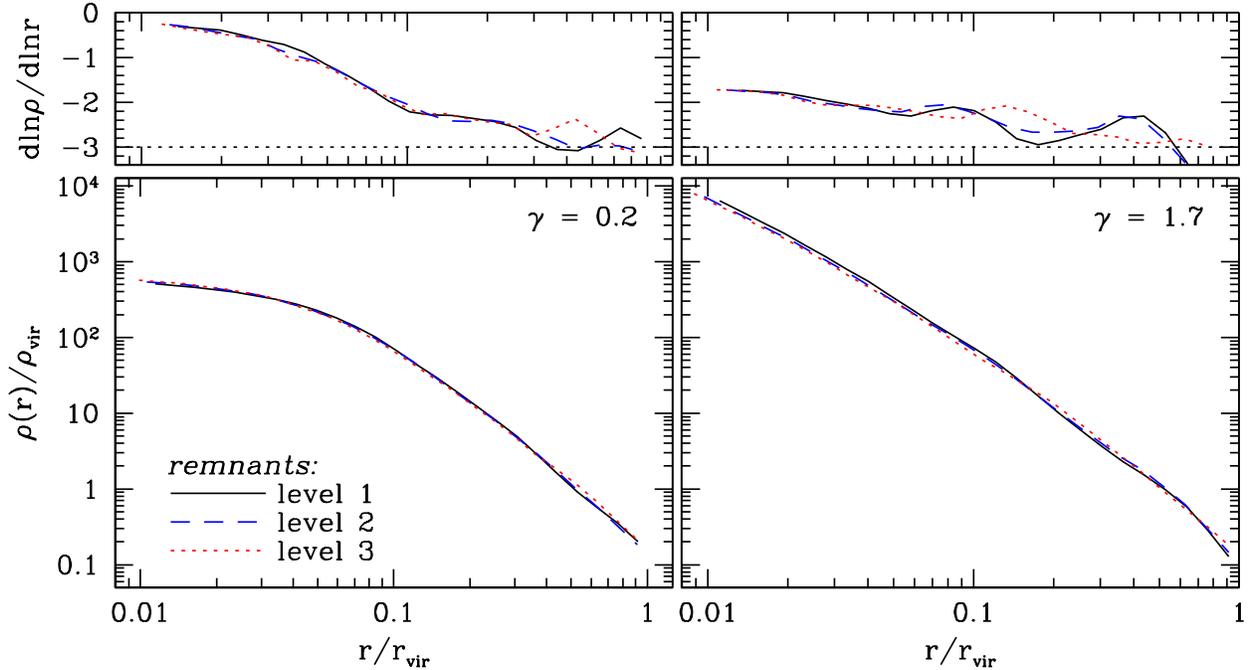}}
\caption{Density and logarithmic slope profiles as in Figure~\ref{fig3}, but 
  for the {\it sequence} of binary, parabolic mergers between identical halos (runs Sp1-Sp6).
  Each merger remnant was used as a progenitor in the next level of mergers 
  after removing all unbound particles and randomly 
  orienting its axis. The asymptotic density slope of 
  the initial profiles is indicated in the 
  {\it upper right-hand corner} of the bottom panels and remnant profiles 
  are normalized to their own virial radii. The profiles 
  from each level of the hierarchy are 
  shown with different line type as indicated in the 
  {\it lower left-hand corner}. Hierarchical merging does not serve to modify the inner 
  slope and overall density profile shape down to the limit of our force resolution.
\label{fig7}}
\end{figure*}

In all sequences of mergers, we identified a time after which the 
central density profile of the remnant did not evolve significantly 
(as before changes of the order of few percent were considered acceptable).
We further allowed the remnants to settle into equilibrium for a timescale 
equal to one crossing time at the virial radius of the system, $t_{\rm cross}(r_{\rm vir})$.
Then their equilibrium state was analyzed. The density profiles 
and logarithmic slopes for two of the remnant hierarchies, starting from the 
shallow and steep initial profiles, are shown in Figure~\ref{fig7}. 
In order to reduce the noise in the measurement, we superposed the outputs 
from $25$ snapshots between the time at which the merger was completed $t_{\rm comp}$, 
and $t_{\rm comp} + t_{\rm cross}(r_{\rm vir})$. Not surprisingly, the 
remnants at all levels of the hierarchy exhibit faithful 
memories of the inner power-law indices of their 
progenitors. This indicates that, even in a complex hierarchy of halo mergers, the inner 
slopes of the density profiles are well preserved. Repeated dissipationless equal-mass 
merging cannot modify the inner power-law index or even the general shape of 
the spherically-averaged density profiles. The evolution of the virial mass in these 
simulations of hierarchical merging is even stronger than the one portrayed in the binary
encounters. Indeed, the fraction of particles found outside of what is 
formally identified as $r_{\rm rvir}$ rises to $\sim 50\%$ 
in the third level of mergers, emphasizing 
the need for more reliable accounting in semi-analytic models. 

\begin{figure}[t]
\centerline{\epsfysize=7.5truein \epsffile{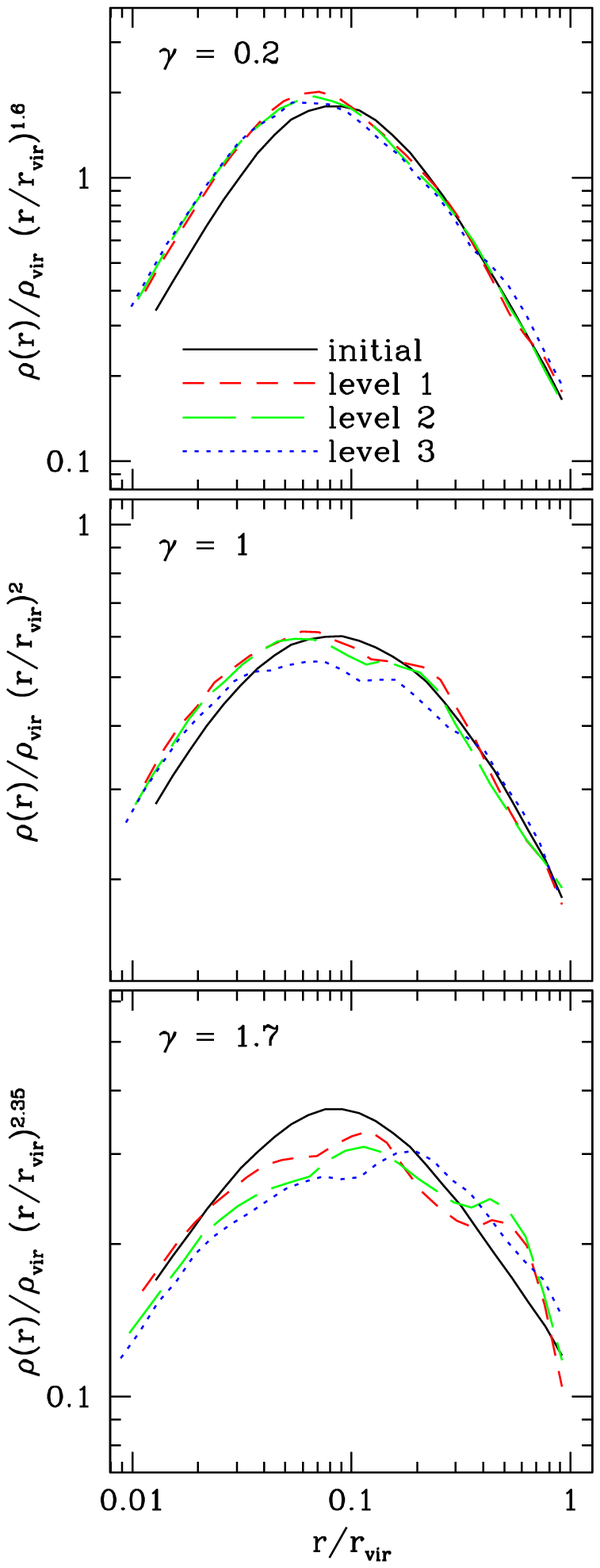}}
\caption{Density profile shapes, scale radii, and concentrations of 
  the remnants in the sequence of binary mergers (runs Sp1-Sp6).
  The central density slopes of the initial profiles 
  are indicated in the {\it upper left corners} of each panel. In order to emphasize 
  the details of the profiles and the scale radius of the transition between the inner and 
  outer power laws, we plot the product $\rho(r) r^{(\beta + \gamma)/2}$ 
  on the vertical axis rather than $\rho(r)$. In each case, 
  the remnant profiles are normalized 
  to their own $r_{\rm vir}$ as defined in \S~\ref{sub:halomodels}. 
  Under the assumption that 
  each remnant can be described by the same set of $[ \alpha, \beta, \gamma ]$ slopes 
  as the initial models, which seems well justified in this case, the maximum of the product 
  $\rho(r) r^{(\beta + \gamma)/2}$ indicates the scale radius of the system.
\label{fig8}}
\end{figure}

Following up on the evolution of the profile shapes, 
an intriguing result of the sequential merger experiments 
is reported in Figure~\ref{fig8}. Here, we 
plot the product $r^{(\beta+\gamma)/2}\rho(r)$, where radii and densities
are normalized to the virial values, as a function of $r/r_{\rm vir}$. The quantity 
$(\beta+\gamma)/2$ is equal to $1.6$, $2.0$, and 
$2.35$ for the shallow, NFW, and steep profiles respectively. 
There are several reasons to plot this quantity. First,
it scales out the gross dependence of the halo profile on 
radius so that small changes become more apparent. Second, the 
radius at which this quantity is maximized corresponds to the scale 
radius, $r_{\rm s}$, of the density distribution. 
The three panels of Figure~\ref{fig8} show that the 
general shape of the remnant density profiles changes 
minimally throughout the merging sequence. In particular, 
the scale radius appears to remain fixed at a nearly 
{\it constant fraction} of the virial radius of the 
halo at all stages. In other words, the concentration parameter remains 
approximately constant during periods of equal-mass merging. 

The strongest evolution appears to be between 
the initial and the level one merger remnant profiles. 
However, this shift is small and is not seen in the 
subsequent mergers in the sequence. We attribute the difference in the 
behavior between the initial model and level one merger remnant to the 
fact that the initial merger involves spherically-symmetric systems, while 
the higher level mergers are between triaxial halos.

Furthermore, it is worth pointing out the presence of wave-like features in
the density profiles at $r \gtrsim 0.2r_{\rm vir}$. 
These features are apparent in Figure~\ref{fig8} as we plot 
a quantity that emphasizes the details of the profiles, though 
they are also noticeable in Figs.~\ref{fig3}, \ref{fig5}, and \ref{fig7}.
They are associated with ``shells'' of particles left over
after the merger and do not disappear even after several billion years of 
evolution subsequent to the completion of the collision. Therefore, even though 
the mergers involve smooth, structureless halos, the density profiles of the remnants 
are not completely smooth.

\begin{figure*}[t]
\centerline{\epsfysize=4truein \epsffile{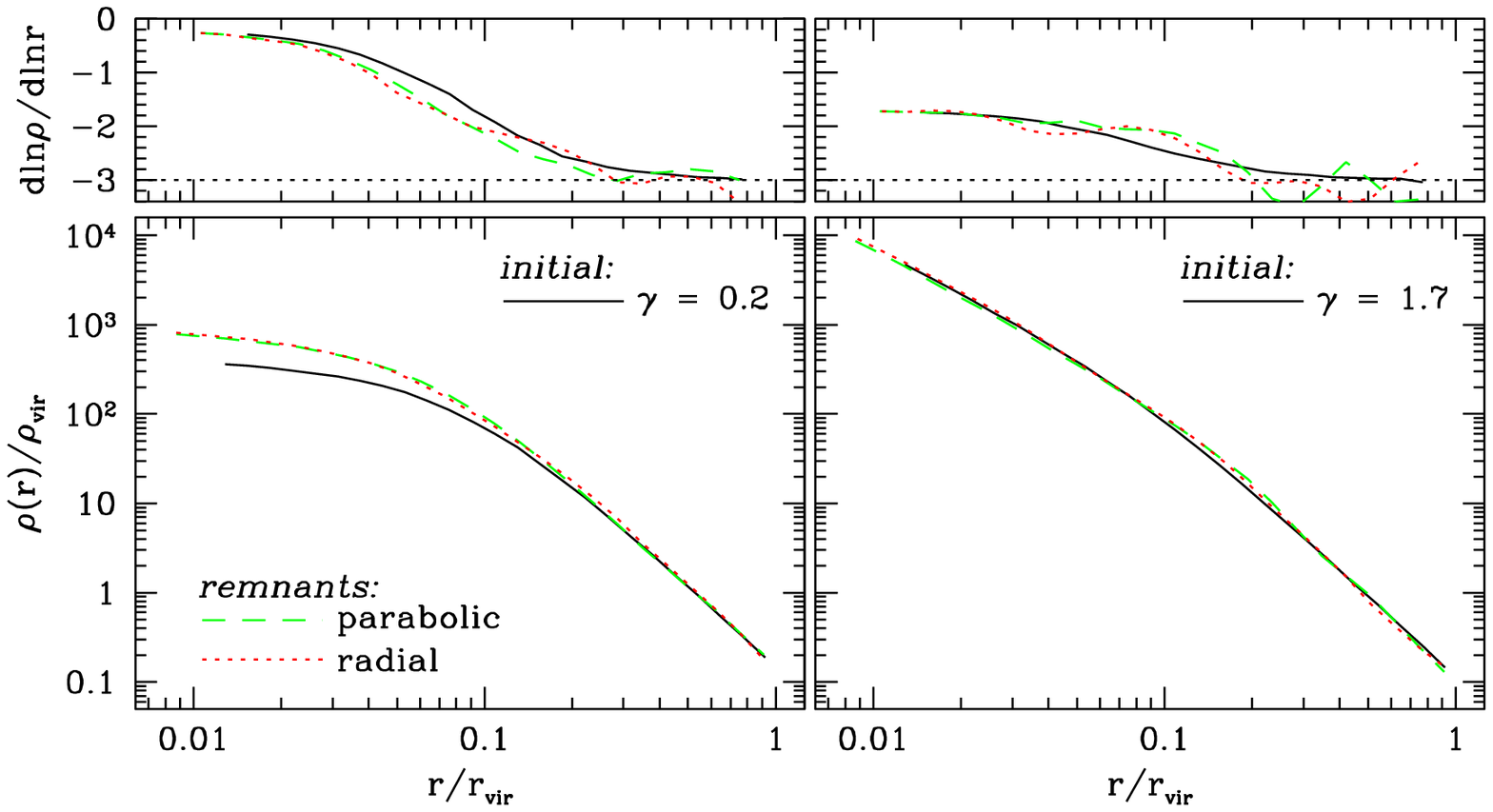}}
\caption{Density and logarithmic slope profiles as in Figure~\ref{fig3}, 
  but for the {\it quadruple} mergers between identical halos  (runs Qp1, Qp2, Qr1 and Qr2). 
  The {\it left-hand panels} show results for the shallow initial density profiles while the 
  {\it right-hand panels} show results for the steep initial density profiles. We plot results for both 
  the parabolic ({\it dashed} lines) and radial ({\it dotted} lines) initial orbital configurations.
\label{fig9}}
\end{figure*}

The last of the DM halo-only merger experiments that we explored
involved the simultaneous collision of four identical halos (runs Qp1, Qp2, Qr1 
and Qr2). We performed these multiple merger experiments for shallow and steep
initial profiles on both parabolic and radial initial orbits as
described in detail in \S~\ref{sub:sims}. One might anticipate that in
such simultaneous multiple mergers potential fluctuations are stronger
and more violent, which may lead to a larger degree of relaxation
compared to the binary mergers. The resulting profiles of $\rho(r)$
and $d \ln \rho(r)/d \ln r$ are shown in Figure~\ref{fig9}. In this
case, again, the merger remnants reflect the inner power-law 
indices and overall profile 
shapes of their progenitors and exhibit, at most, 
only a small shift in their scale radii relative to 
their virial radii. This reinforces our basic result, 
that the density structures of the merger remnants are nearly the 
same as their progenitors. This result cannot be
circumvented by appealing to a complex merger hierarchy or episodes of
multiple, nearly simultaneous, equal-mass mergers. In all cases, the
final systems exhibit a remarkable memory of the density structure of their 
progenitors.

\subsection{Galaxy Mergers}
\label{section:galaxy_mergers}

One of the goals of this study is to assess the general applicability 
and robustness of our results and the results of previous studies 
regarding the evolution of density profiles during dissipationless 
equal-mass mergers. As a further investigation of the generality of 
our primary findings, we perform numerical simulations of binary 
mergers between identical disk and elliptical galaxies constructed to
resemble present-day systems. The models we adopt do not include 
gaseous dissipation, but include the various dissipationless components thought 
to dominate the mass distribution of real galaxies as well as 
adiabatic contraction of the DM halo during galaxy formation. 
These additional features allow orbital energy to be 
dissipated more quickly, the merger to come to completion significantly sooner, 
and the merger remnants to have different shapes with respect to the
DM halo-only mergers \citep{kazantzidis_etal04c}. 

\begin{figure*}[t]
\centerline{\epsfysize=4truein \epsffile{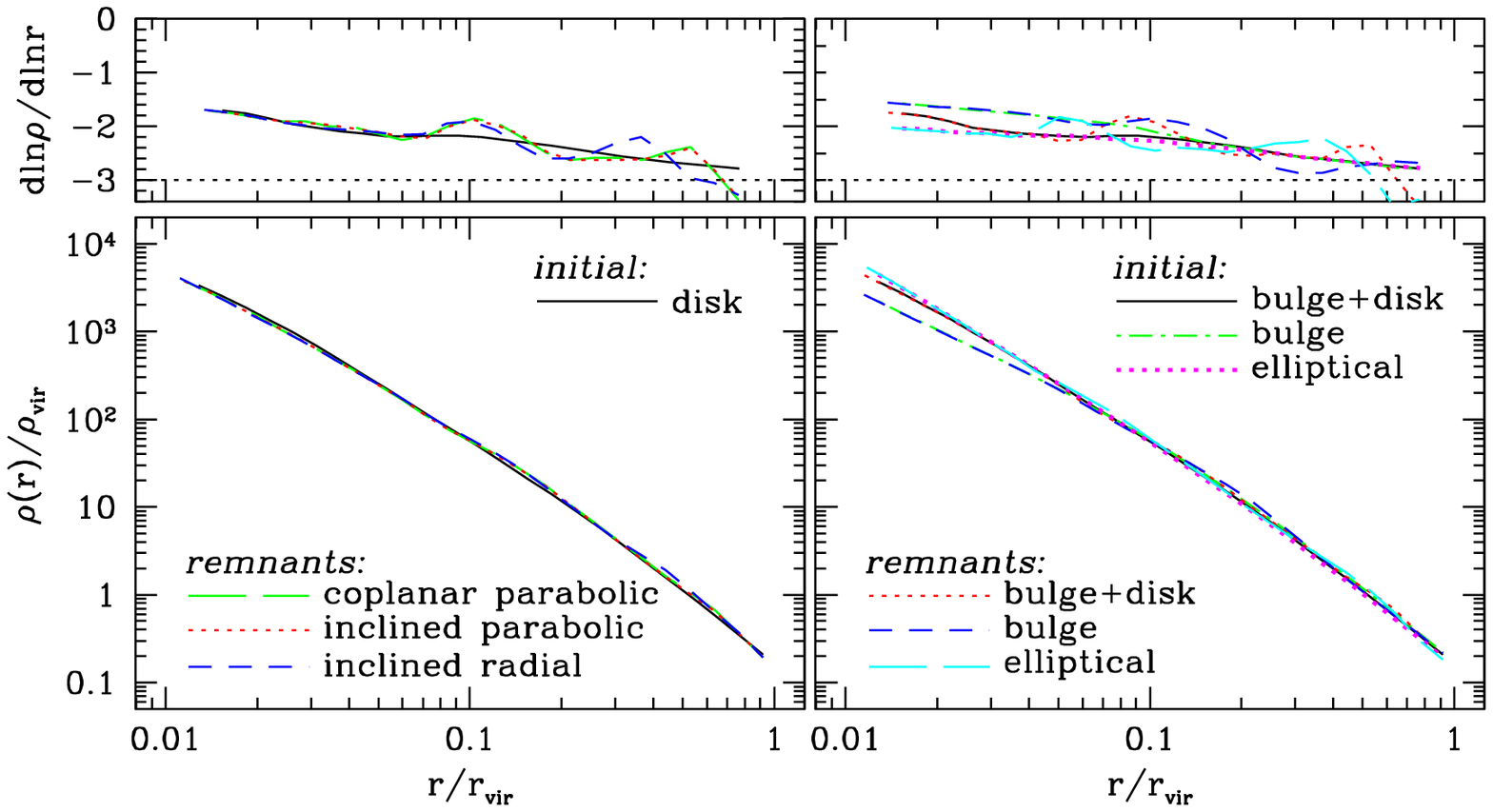}}
\caption{{\it Left:} Density and logarithmic slope profiles of {\it DM} as in 
  Figure~\ref{fig3}, but for the binary merger simulations between identical 
  halo$+$disk {\it galaxy} models (runs dgBp1, dgBp2 and dgBr). The initial profile 
  is shown by the {\it solid} line. Different line types correspond to various
  orbital configurations and relative disk inclinations as indicated in the 
  {\it lower left-hand corner} of the bottom panel.
  {\it Right:} Profiles of {\it DM} in binary, parabolic merger 
  simulations between identical multicomponent systems. 
  Results are reported for systems consisting of a DM halo, 
  a stellar disk, and a compact bulge (run hbdBp), systems comprised of a DM 
  halo and a bulge (run hbBp), and elliptical galaxies (run egBp). The presence of cold 
  baryonic components preserves the density structure of DM in mergers between identical 
  systems. 
\label{fig10}}
\end{figure*}

In the disk galaxy mergers (runs dgBp1, dgBp2, dgBr and hbdBp), two identical spirals with
self-gravitating disks, rotating DM halos and optional compact bulges merge in
a variety of initial orbital configurations and relative
orientations of the stellar disks. In the halo$+$bulge (run hbBp) and elliptical galaxy merger 
(run egBp), two identical systems with non-rotating DM halos and spherical stellar
distributions collide on a parabolic orbit. Figure~\ref{fig10} displays 
results for all galaxy mergers performed in this study. In the left panels 
of Figure~\ref{fig10}, we show the density and logarithmic slope profiles 
of {\it DM} in the halo$+$disk mergers with no stellar bulge
component. The results concerning all other galaxy merger simulations
are presented in the right panel of Figure~\ref{fig10}. 

The central density slopes and the overall density profile
shapes of DM are maintained in mergers between identical systems consisting 
of multiple collisionless components. Hence, the presence of cold baryonic
components does not render the chief result of this inquiry, 
that dissipationless mergers result 
in remnants that are practically scaled versions of their
progenitors, ineffectual. \citet{boylan-kolchin_etal05} reached a similar 
conclusion in dissipationless mergers of elliptical galaxies for several different 
orbital configurations and mass models. The overall significance of our findings
lies in the fact that the mixing of the various collisionless components during the 
mergers occurs in a manner that preserves the overall density structure of the 
DM component.

%

\section{Discussion}
\label{sec:discussion}

The results presented in this paper have interesting implications for
the formation of a universal density profile in cosmological $N$-body
simulations.  We have established that the characteristic power-law
indices of halos are not affected by major mergers, which should
be ubiquitous during the early rapid-mass-accretion stages of halo
evolution. Instead, this universal profile shape must be something
that is imprinted on halos at the time of the collapse of the first
gravitationally-bound objects. In hierarchical cosmological models,
the amount of power in the initial spectrum of density perturbations
is typically washed out on scales smaller than some critical scale.
For example, in the popular case of the lightest supersymmetric
particle this scale is set by the kinetic energies of the DM particles
when they kinetically decouple from the thermal fluid
\citep[e.g.,][]{chen_etal01,hofmann_etal01}.  Thus, there exists a
typical ``smallest'' mass for collapsed structures, which is of order
of $\sim 10^{-7}$--$10^{-5}\ M_{\odot}$ for the most frequently
studied model of supersymmetric neutralino DM. The results reported in
this paper can be used to argue that these first smallest structures
must already have density profiles that follow an NFW-like universal
form.  This conclusion is in agreement with findings of direct
numerical simulations. For example, \citet{huss_etal99} and
\citet{shapiro_etal04} find that gravitational collapse leads to
NFW-like profiles for a wide variety of initial conditions, while
\citet{moore_etal99}and \citet{diemand_etal05} show that in
cosmological $N$-body simulations of the first virialized objects with
masses near truncation mass of the power spectrum, these early halos
have NFW-like density profiles.

Given that dissipationless late-time major mergers do not modify the shape
of the profile, the problem of understanding the universality of the
NFW-like density distribution can be relegated to understanding the
reason why initial, rapid gravitational collapse leads to a density
structure that is close to NFW. However, this is likely to be only an 
approximate picture. The actual mass assembly involves
minor mergers, for which the density profile of the remnant
will be a non-trivial average of the profiles of the progenitors, as
smaller halos sink into the center and modify the inner density
distribution \citep[e.g.,][]{syer_white98,dekel_etal03,ma_boylan-kolchin04}.
Therefore, it may be expected that the details of merger histories
will induce some evolution of profile shape \citep[e.g.,][]{gao_etal05},
which may explain the existence of significant scatter in the inner
density profiles of cosmological structures \citep{tasitsiomi_etal04}.
Nevertheless, in the early rapid-accretion stage of halo evolution,
the fraction of mass contributed by minor mergers should be relatively
small and the bulk of the mass should be assembled via major mergers
\citep[e.g.,][]{lacey_cole93,lacey_cole94,zentner_bullock03} so the
above conclusion can still be applicable.

In the context of the scaling relations of DM halos in cosmological
$N$-body simulations, another intriguing result is reported in
Figure~\ref{fig8}.  This plot illustrates that halo concentrations
remain approximately constant through a sequence of mergers between
identical systems. The controlled simulations we presented do not
include the constant redefinition of the virial radii of {\it
  individual systems} that is due to the dilution of the mean
background density with redshift.  This dilution leads to a definition
of virial radius that grows as $r_{\rm
  vir}(a) \propto a$, where $a$ denotes the scale factor. Combining
this growth of virial radii with our measurement of the constancy of
$c$, leads to a scaling of concentration as
$c \propto r_{\rm vir}(a)/r_{\rm s} \propto a$ during cosmological
evolution. This scaling is in broad agreement with the findings of the
redshift evolution of halo concentrations in cosmological $N$-body
simulations subsequent to an early phase of rapid-mass-accretion 
\citep[e.g.,][]{bullock_etal01a,wechsler_etal02,zhao_etal03,
  tasitsiomi_etal04,dolag_etal04}. An interesting deviation from this
is that \citet{zhao_etal03} found
that during the early rapid-mass-accretion period of halo growth, halo
concentrations appear to remain approximately constant, suggesting
qualitatively different behavior during the first stages of the
collapse of rare density peaks.

We assessed the generality of our 
conclusions by initializing a sequence of parabolic 
encounters according to the conditions of 
cosmological mergers at redshift $z=3$, rather than at $z=0$.  
The initial systems followed the steep density profile and had 
the same $M_{\rm vir}$ as the standard halo models. We scaled the 
concentration parameter according to $c \propto (1+z)^{-1}$ for fixed mass objects 
\citep{bullock_etal01a,wechsler_etal02,zhao_etal03} 
and assumed that the first level of mergers initiates at $z=3$.  
Our modeling results in a substantially denser initial system 
than before with $r_{\rm vir} \simeq 79.5 \kpc$, 
and correspondingly smaller orbital times. 
Indeed, the first level of the hierarchy 
concludes after only $\sim 1.5$ Gyr, 
as opposed to $\sim 5$~Gyr for the $z=0$ case.  
The second and third level of mergers complete after
$\sim 1.8$~Gyr and $\sim 2.1$~Gyr, respectively. 
The remnants at all levels of the high-z merger 
sequence retain excellent memories of the inner power-law 
indices and density profile shapes of their progenitors, 
in accord with the merger sequences of our
fiducial simulations. Furthermore, the fraction of particles outside 
the virial radii of the remnants at each level is 
substantial (between $\sim 30$ and $\sim 50\%$),
in agreement with the findings reported in \S~\ref{sub:multi_halo}.
As expected, the results of our study are insensitive to the overall 
densities and merger timescales that characterize mergers occuring at 
different cosmological epochs.

Recently, \citet{lu_etal05} have argued that frequent major mergers
associated with the rapid-mass-accretion phase may set the inner
$\rho\propto r^{-1}$ profiles of CDM halos by effectively isotropizing
the velocity field of DM particles during violent relaxation.  
Our simulations show that the shape of the density profile is well preserved 
during equal-mass mergers. Violent relaxation is therefore 
{\it incomplete}, as it does not erase memory of the progenitor properties. Further 
evidence to this is provided by numerical studies of mergers showing that the
potential fluctuations necessary for violent relaxation are damped
before the binding energies of particles are completely randomized
\citep[e.g.,][]{white78,barnes92}. We also observe that particle
orbits in merger remnants, especially for the initial models with
steep inner cusps, exhibit radial anisotropy at all radii. Although
the initial progenitor models are fully isotropic (i.e., $\beta\equiv
1-\sigma_{\rm t}^2/2\sigma_r^2=0$ at all radii, where $\sigma_{\rm t}$
and $\sigma_r$ are the tangential and radial velocity dispersion, respectively), 
merger remnants exhibit anisotropy that changes from mildly radial with 
$\beta\approx 0-0.2$ at small radii ($r/r_{\rm vir}\lesssim 0.1$) to strongly
radial with $\beta\approx 0.3-0.4$ on the outskirts \citep[see also][]{boylan-kolchin_ma04,
moore_etal04,dekel_etal05}. Such a radial dependence of the shape of
$\beta$ is rather similar to that observed in CDM halos formed 
in cosmological simulations \citep[e.g.,][]{cole_lacey96,colin_etal00,faltenbacher_etal05}.  
Therefore, equal-mass mergers do not lead to isotropic orbits. Nonetheless,  
we have probed the effect of mergers in the specific regime of 
equal-mass encounters that take place over several dynamical times.
The situation may be qualitatively different when 
a large number of mergers, occuring in rapid succession, 
dramatically increase the depth of the potential over dynamical timescales, 
as assumed by \citet{lu_etal05}. Such rapid evolution of potential is likely
to be relevant for the very earliest phases of collapse.  

Our findings can be used to place constraints on the expectations for
the DM distribution in galaxies and clusters. \citet{loeb_peebles03}
and \citet{gao_etal04} have put forward the hypothesis that one of the
fundamental properties of the NFW density profile is that it is a
dynamical attractor in collisionless evolution. More specifically,
these authors argued that if the density structure of a collisionless
system were to deviate from the NFW profile, perhaps due to baryonic
dissipation or some other process, subsequent dissipationless major
merging would drive the density distribution toward the NFW form. This
``attractor'' hypothesis, combined with observations that stars
dominate gravitationally in the centers of galaxies, would imply
shallower DM density profiles in the inner regions of clusters than
predicted by collisionless cosmological simulations
\citep[e.g.,][]{sand_etal02}. However, the numerical results presented
in this study do not support the attractor hypothesis. Our simulations
show that the inner power-law indices and the overall shapes of 
density profiles remain unmodified during collisionless equal-mass binary
mergers, even after a sequence of collisions, or in simultaneous
multiple mergers. This conclusion persists even if the progenitor
systems contain cold stellar components with a variety of properties.

An immediate consequence is that the effects of gas cooling 
and baryon condensation in halo centers, which steepen the inner DM 
density profiles, are maintained in subsequent dissipationless major merging. 
This is consistent with the results of \citet{gnedin_etal04}, who showed that the 
effects of cooling at high redshifts are retained in cosmological simulations to the present day 
and can be well described by the adiabatic contraction model, 
even when DM halos undergo a period of dissipationless evolution.

One of the many routes to forming bright elliptical galaxies in hierarchical
cosmological models involves dissipationless merging in their late stages of evolution.
Though the details of this picture have yet to come into focus, the results 
reported in this study are relevant to scenarios of elliptical galaxy formation. 
Observed bright and faint ellipticals exhibit a dichotomy in their inner stellar
density distributions. Bright ellipticals tend to have flatter central
stellar density profiles compared to faint ellipticals, which exhibit
cuspy profiles \citep[e.g.,][]{faber_etal97,kormendy99}. Our findings
intimate that dissipationless hierarchical merging of faint ellipticals alone
cannot alone be invoked as a mechanism for forming bright elliptical 
galaxies. Any such scenario would require an additional process that should be
responsible for the flattening of the inner density distributions. As
an example, the prevalence of relatively low-density cores in bright
elliptical galaxies may be a consequence of the existence of
supermassive black holes and their interactions with the stellar backgrounds 
\citep{makino_etal96,merritt_cruz01}.

An analytical framework exists that aids in the understanding of our numerical results. 
Undoubtedly, our dissipationless simulations must conform to Liouville's theorem 
demanding that the phase-space density of a dynamical system that undergoes 
collisionless evolution to a new state cannot increase.
Nevertheless, as has also been emphasized by \citet{boylan-kolchin_ma04}, 
Liouville's theorem applies to a six-dimensional quantity and 
there is no {\it a priori} reason to expect that both configuration 
and momentum space distributions should be independently conserved.  
In general the distributions in these subspaces are {\it not} conserved.  
In fact, our simulations show that both configuration and momentum 
space distributions are altered during mergers.

More complicated treatment of phase-space evolution is needed to draw
conclusions about evolution of density profiles \citep{mathur88,dehnen05}.
In particular, \citet{dehnen05} recently showed that for
phase-space density $f$, the {\it excess mass function} $D(f)$,
defined as the difference between the mass with phase-space density 
$>f$ and the product of $f$ and the volume of phase-space with density $>f$, 
can only decrease upon mixing. This mixing theorem and the simple properties 
of $D(f)$ as $f \rightarrow \infty$ for self-gravitating density cusps can then
be used to prove that a merger remnant cannot have a density cusp
steeper than that of the steepest progenitor. \citet{dehnen05} also
argued that reducing the slope of the density cusp would require
arbitrarily large dilution of the phase-space density so that it seems
implausible that the remnant could have an inner cusp shallower than
the steepest progenitor cusp.  

Our findings are in agreement with the
conclusions of \citet{dehnen05}. Indeed, Figure~\ref{fig3} illustrates that in a 
merger of two identical power-law density profiles, the 
remnant exhibits a power-law inner density slope equal to that of the progenitors. 
In addition, the results embodied by Figure~\ref{fig5} hint that in mergers between 
halos of very different inner density power laws, the remnant has a central density slope slightly 
shallower than that of the steepest of the progenitors on scales that are of practical concern 
for galaxies. However, the results of \citet{dehnen05} apply asymptotically as $r \rightarrow 0$ 
and our simulations simply may not resolve sufficiently small radii for 
the asymptotic result to hold. In this regard, we see no evidence that the 
second conclusion of \citet{dehnen05} is violated but higher-resolution numerical 
simulations of merging density cusps are required to test this conjecture more extensively. 

Our findings also shed light on the resilience of density cusps to 
gravitational interactions. \citet{kazantzidis_etal04b} elucidated the dynamical effect 
of tides on the central density profiles of cuspy satellite halos.  
These authors employed collisionless, high-resolution $N$-body cosmological simulations 
with simulations of the tidal stripping of individual 
satellites orbiting within a static host 
potential and showed that the density distribution retains 
its initial inner power-law index as the 
satellite experiences mass loss. Their analysis revealed that the central density cusp is 
notably stable to tidal shocks and gravitational stripping. The results reported in 
this paper combined with those of \citet{kazantzidis_etal04b}, suggest that density cusps, 
once formed, are extremely difficult to disrupt or even to modify in any significant way.  

The preservation of the inner power laws of halo density profiles has
intriguing implications for tests of the nature of DM through
observations of galactic and sub-galactic structure.  In alternatives
to CDM models such as WDM
\citep[e.g.,][]{pagels_primack82,colombi_etal96,hogan_dalcanton00} or
DDM \citep[e.g.,][]{kaplinghat05,cembranos_etal05} the DM particles
typically have non-negligible momenta which leads to a suppression of
small-scale structure for two reasons. The first reason is that the
power spectrum of density fluctuations on small scales is damped by
the free-streaming of the particles
\citep[e.g.,][]{ma96,bode_etal01,kaplinghat05}.The second reason is
that these particles are restricted to finite phase-space densities
\citep{tremaine_gunn79,hogan_dalcanton00,kaplinghat05}. The
phase-space constraint is thought to induce a cored inner profile of
nearly constant density in small and early-forming DM halos, because
the very high phase-space densities of cusp-like profiles are
unachievable \citep[e.g.,][]{avila-reese_etal01}.  The findings of the
present study imply that if the small, early-forming building blocks
of larger halos were formed with cored density profiles, these cores
should be well-preserved during the hierarchical assembly of larger
halos, in the very least during periods of rapid major mergers.  The
signatures of the alternative DM could, therefore, be possibly
observed in accurate measurements of the rotation curves of
DM-dominated galaxies at small scales
\citep[e.g.,][]{deblok_etal01,salucci_etal03,weldrake_etal03,simon_etal04}
or the flux anomalies in strong-lens systems
\citep[e.g.,][]{dalal_kochanek02,zentner_bullock03,rozo_etal05}. This
could potentially provide fundamental constraints on the properties of
the DM particles.

We conclude by emphasizing that in this article, 
we have only addressed the evolution of 
density profiles in mergers between equal-mass systems.  
Recent works have examined more 
general situations involving systems with a spectrum of 
masses and reached similar conclusions \citep{boylan-kolchin_ma04,aceves_velazquez05}.
This fact allows us to argue that our main findings can be generalized
to all unequal-mass {\it major} mergers with some degree of
confidence.

In our modeling, we have also implicitly assumed that cuspy halos have
a higher phase-space density in the inner regions than cored profiles.
The fact that in hybrid mergers the inner density slope of the remnant
is closer to that of the steepest of the progenitors suggests that the
quantity that drives the evolution of the profile is the phase-space
density. However, it is not difficult to imagine merging scenarios
where this situation can be reversed as, for example, in cases of
unequal-mass mergers or encounters between systems with considerably
different concentration parameters. We defer a detailed numerical
study of these considerations to future work.

\section{Summary}
\label{sec:summary}

Using a large suite of controlled, dissipationless $N$-body simulations 
we have explored the evolution of DM density profiles in equal-mass mergers of 
DM halos and multicomponent galaxies. A common 
feature of the majority of previous related investigations is that they 
did not consider the effect of hierarchical merging that characterizes 
CDM models, nor did they examine the role of 
baryonic components on the density structure of merger remnants.
In contrast, our numerical experiments are performed in degrees of
increasing complexity. We present an ensemble of simulations comprised not only 
of binary encounters, but also sequences of mergers and collisions between
multiple progenitors in order to study the range of behaviors
that are realized in the context of cosmological structure formation.
We have also probed a wider range of parameter space with respect to earlier
studies by varying the initial orbital configurations and orbital energies 
in an effort to establish the generality of our main results.
Finally, we examined collisions between systems with 
multiple components consisting of rotating DM halos, 
stellar disks, and stellar spheroids. This enabled us to elucidate the impact of 
internal angular momentum and the presence of cold baryons on our findings.
We find that the merger remnants in our simulations 
conform to a set of several simple and general 
results that can be summarized as follows.

\begin{itemize}

\item[1.] Binary mergers between identical DM halos with density structures 
  described by various asymptotic power-law indices $\rho \propto r^{-\gamma}$ 
  ranging from steep cusps to core-like profiles produce remnants with inner 
  density slopes equal to those of the progenitors.  
  Furthermore, the overall shape of the remnant 
  density distribution constitutes a remarkable reflection of that of the initial systems.  
  If the progenitor halos are constructed with appreciably different asymptotic 
  power-law indices, the inner density slope of the remnant 
  is closer to that of the steepest of the initial systems.  These conclusions hold for a 
  wide range of orbital energies. 

\item[2.] The aforementioned findings remain valid for a variety of encounter configurations 
  including sequences of several consecutive merger events, designed to mimic hierarchical 
  merging, and simultaneous collisions of several systems.  Overall, the inner slopes and 
  shapes of density profiles are remarkably robust during dissipationless evolution, 
  regardless of the number of mergers and initial conditions associated with the encounters.
  Dissipationless equal-mass mergers do not provide a mechanism for 
  driving the density profiles of collisionless matter toward a universal form,
  contradicting the hypothesis that the NFW profile behaves as a dynamical 
  attractor \citep{loeb_peebles03,gao_etal04}.
  
\item[3.] In binary mergers between identical systems, particles from
  both systems contribute equally to the final density profile at all
  radii, as must be the case due to symmetry. In mergers between systems
  of markedly different inner density power laws, the central regions of the
  remnants are dominated by particles originating from the system with
  the steeper inner density slope. The steepest inner core survives,
  almost unaffected, in the remnant.

\item[4.] Remnants in mergers between equal-mass systems 
  contain significant fractions of their
  bound mass, $\sim 25-50\%$, well beyond their formal virial radii and extending 
  out to $\approx 2-4 r_{\rm vir}$. This suggests that the virial 
  mass is not simply additive in mergers, as is commonly assumed in 
  many semi-analytic models of halo and galaxy evolution. Semi-analytic models 
  may significantly overestimate the mass and density 
  within the virial radii of merger remnants.

\item[5.] Mergers between identical DM halos produce self-similar
  remnants in the sense that their scale radii, $r_{\rm s}$, defined as
  the radii at which the density profiles transition from an inner to
  an outer power law, are very similar to those of the progenitors when
  measured in units of the halo virial radii, $r_{\rm vir}$. In other
  words, halo concentrations remain essentially unmodified during equal-mass 
  encounters and the remnants correspond to scaled versions of their progenitors.

\item[6.] Mergers between identical systems containing cold stellar components in the form of 
  disks, spheroids, or a combination of the two, in addition to the extended DM halos, 
  exhibit a degree of self-similarity akin to those of the DM halo-only mergers. Namely, the 
  overall shape of the density profiles and the inner density slopes of DM are maintained in the 
  remnants. Particle mixing preserves the density structure of the DM component. 
  The effects of gaseous dissipation on the DM profiles of 
  the progenitor systems will be retained in the density 
  distribution of their descendants under dissipationless major merging.

\end{itemize}

\acknowledgments

The authors acknowledge many stimulating discussions with Tom Abel, Michael
Boylan-Kolchin, Avishai Dekel, Vincent Eke, Oleg Gnedin, Anatoly Klypin, Abraham Loeb, 
John Magorrian, Gary Mamon, Lucio Mayer, David Merritt, Houjun Mo, 
Daisuke Nagai, Robert Sakamano, Joachim Stadel, and Risa Wechsler. 
SK is supported by the Swiss National Science Foundation and by The Kavli Institute for 
Cosmological Physics (KICP) at The University of 
Chicago. ARZ is funded by the KICP and the National Science Foundation (NSF) under grant 
No. NSF PHY 0114422. AVK is supported by the NSF under grants No. AST-0206216 and
AST-0239759, by NASA through grant NAG5-13274, and by the KICP. The
numerical simulations used in this study were performed on the zBox
supercomputer at The University of Z\"urich and on the Intel cluster
at the Cineca Supercomputing Center in Bologna. This research made use of the NASA 
Astrophysics Data System.

\bibliography{profiles}

\end{document}